\documentclass{aa}  

\usepackage{graphicx}
\usepackage{txfonts}
\usepackage[dvipsnames]{xcolor}
\usepackage{ulem}
\usepackage{nicefrac}
\usepackage{tcolorbox}
\usepackage{hyperref}
\usepackage[utf8]{inputenc}
\usepackage{verbatim}
\hypersetup{
    colorlinks=true,
    citecolor=blue
}

\newcommand{\GG}[1]{}

\newcommand{\pa}{\psi_\textrm{CZ}}
\newcommand{\pb}{\varepsilon}

\newcommand{\spec}{N_\textrm{sp}}

\begin{document} 

    \title{Properties of the ionisation glitch}
    \subtitle{I. Modelling the ionisation region}

   \author{Pierre S. Houdayer \inst{1} \and Daniel R. Reese \inst{1} \and Marie-Jo Goupil \inst{1} \and Yveline Lebreton \inst{1,2}}

   \institute{LESIA, Observatoire de Paris, Université PSL, CNRS, Sorbonne Université, Université de Paris, 5 place Jules Janssen, 92195 Meudon, France 
   \and Univ Rennes, CNRS, IPR (Institut de Physique de Rennes) – UMR 6251, 35000 Rennes, France}

   \date{Received 5 July 2021 / Accepted 30 September 2021}

 
  \abstract
   {Determining the properties of solar-like oscillating stars can be subject to many biases. A particularly important example is the helium-mass degeneracy, where the uncertainties regarding the internal physics can cause a poor determination of both the mass and surface helium content. Accordingly, an independent helium estimate is needed to overcome this degeneracy. A promising way to obtain such an estimate is to exploit the so-called ionisation glitch, i.e. a deviation from the asymptotic oscillation frequency pattern caused by the rapid structural variation in the He ionisation zones.}
   {Although progressively becoming more sophisticated, the glitch-based approach faces problems inherent to its current modelling such as the need for calibration by realistic stellar models. This requires a physical model of the ionisation region explicitly involving the parameters of interest such as the surface helium abundance, $Y_s$. }
   {Through a thermodynamic treatment of the ionisation region, an analytical approximation for the first adiabatic exponent $\Gamma_1$ is presented.}
   {The induced stellar structure is found to depend on only three parameters including the surface helium abundance $Y_s$ and the electron degeneracy $\psi_\textrm{CZ}$ in the convective region. The model thus defined allows a wide variety of structures to be described and, in particular, is able to approximate a realistic model in the ionisation region. The modelling work conducted enables us to study the structural perturbations causing the glitch. More elaborate forms of perturbations than the ones usually assumed are found. It is also suggested that there might be a stronger dependence of the structure on both the electron degeneracy in the convection zone and on the position of the ionisation region rather than on the amount of helium itself.}
   {When analysing the ionisation glitch signature, we emphasise the importance of having a relationship that can take into account these additional dependencies.}


   \maketitle

\section{Introduction \label{INTRO}}

\begin{figure*}
\centering
\includegraphics[width=\textwidth]{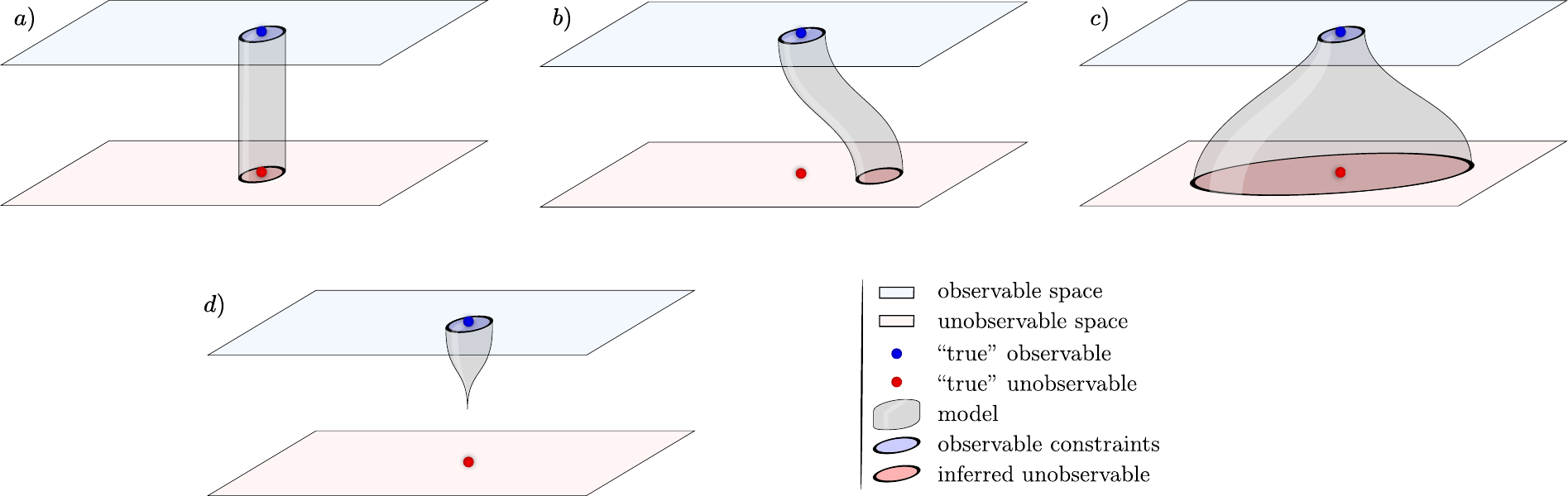}
\caption{Schematic illustration of several potential biases involved in the inference problem. Applied to the abundance determination problem, the observable space represents (notably) the oscillation frequency sets while the unobservable space would represent the model abundances. Case (a) shows an ideal scenario where, for given constraints (blue circle) our modelling (in grey) provides unbiased abundances (red circle). Panel (b) highlights what can be the expression of physical processes missing or poorly taken into account in the model, thus resulting in \textit{physical biases} although having ideal observables. Case (c) provides an example of \textit{degeneracy}, i.e. parameters that can vary widely because they are too little constrained by the observables. The last panel (d) typically illustrates the solar problem, where no mixture seems to satisfy the constraints. In this case, in order to provide a solution, one must consider inconsistencies in the model's physics or much weaker constraints than actually provided by the oscillation frequencies.}
    \label{AllBiases.pdf}%
\end{figure*}

Asteroseismology, i.e. the study of resonant modes in stars, reveals information on the physical properties of the layers that the wave passes through on its way to the surface. Coupled with a physical model of the star, it thus allows us to constrain the various internal processes involved more than any other method. Undoubtedly, the constraints thus obtained depend on what one chooses to model or not and what one assumes to be known or not. This choice is highly complex, and largely dependent on the quality of the information provided (i.e. the precision available on the observables). But, although high-precision photometric data provided by CoRoT \citep{Baglin2006}, \textit{Kepler} \citep{Gilliland2010,Lund2017} and now TESS \citep{Ricker2015,Stassun2019} allows ever more precise estimates of oscillation frequencies, the evaluation of certain stellar parameters remains uncertain. This illustrates that \textit{accuracy}, rather than \textit{precision}, becomes in this case the limiting factor. Indeed, and although the value of seismic inference in the determination of stellar parameters is undeniable, it also opens the door to potential biases on these very parameters (some of them being illustrated in Fig. \ref{AllBiases.pdf}).

In this respect, the determination of abundances using realistic stellar models constitutes a textbook case, the latter being potentially susceptible to a wide range of biases. Identified sources of difficulties undoubtedly reside in the complexity of the physics that must be considered. In particular, one can evoke the equation of state, opacities \citep{Kosovichev1992} or transport processes such as diffusion, turbulent mixing \citep{JCD1993}, or radiative acceleration \citep{Deal2018} whose consideration or not may result in \textit{physical biases}. Recent studies using model grids also suggest that there is a strong anti-correlation between the mass and initial helium abundance estimates \citep{Lebreton2014,Noll2021}. This \textit{degeneracy} makes the inference more complex by resulting in high volatility of both parameter estimates. Moreover, in the case of a large number of frequencies being available like that of the Sun, inversion techniques highlight significant discrepancies and solar models are forced to compromise between inconsistent abundances, densities or convective zone (CZ) depths \citep{Basu2004b,Asplund2009,Serenelli2009}. In addition to these processes directly involving the composition, one must face additional uncertainties surrounding the near-surface region, namely \textit{surface effects} \citep{JCD1988}. Due to theoretical developments \citep{Canuto1997,Belkacem2021} and 3D hydrodynamical simulations \citep{Belkacem2019,Schou2020}, our understanding of the mechanisms involved in these effects (as well as their contributions) has definitely improved. However, the more costly procedures \citep{Mosumgaard2020} are generally avoided in favour of ad-hoc corrections of the oscillation frequencies \citep{Kjeldsen2008,Ball2014,Sonoi2015} which may impact stellar parameter estimates \citep{Nsamba2018}.

Methods allowing independent measurements of abundances (in particular helium) are therefore the subject of much investigation. The challenge is in fact twofold, since a reliable estimate of abundances would in turn provide a constraint on the internal physics that depends on it.\newline

The specific influence between composition and seismic properties of a star is long known and can easily be understood through the structural change caused by ionisation if ever the abundances were to vary. The fact that this change is localised (in this case in the near-surface region) causes what is called a \textit{seismic glitch} and manifests itself through an oscillatory deviation of the observed frequencies from a chosen reference \citep{Gough1990}. Although this effect is not specific to the ionisation regions, the latter have benefited from numerous treatments, being both the most pronounced glitch and a potential marker of the helium abundance \citep{PerezHernandez1994,Lopes1997}. From this point on, many studies \citep{Monteiro1994,Basu1994,Monteiro1998,Monteiro2005,Gough2002,Houdek2007} considered various shapes of structural perturbations to analytically derive the expected frequency shifts. Most procedures exploit inversion formulae \citep{Dziembowski1990,Gough1991,Antia1994} and rely on an analytical modelling of the glitch to give a parameterised form of the oscillation. The expression thus obtained only needs to be fitted to data in order to provide information on the parameters introduced. Also, in order to overcome dependencies on irrelevant disturbances in the frequencies (e.g. surface effects in our case) it has been proposed to study the second differences, $d^2\nu$, rather than the frequencies directly \citep{Gough1990}: 
\begin{equation}
    d^2\nu_{n,\ell} = \nu_{n+1,\ell}-2\nu_{n,\ell}+\nu_{n-1,\ell}
\end{equation} where $n$ and $\ell$ are the associated oscillation radial order and degree respectively. The seismic diagnostic thus defined is less sensitive to surface effects and core perturbations while highlighting signal that would come from an intermediate acoustic depth \citep{Ballot2004}.

The main benefit of this (now usual) approach is to avoid complex modelling and therefore part of the issues mentioned above. Being based on a model with far fewer parameters, which reflects a local rather than global structure, the method seeks to make the best use of the information residing only in the low degree frequencies. Additionally, this procedure is lighter than a direct minimisation or inversion of the frequency differences thus making it convenient to apply to large samples of stars \citep{Verma2019}. On the other hand, it would be a mistake to think that the procedure is ``model independent'' as it depends on the form of the perturbation considered. Therefore, such a procedure may well lead to nonphysical or inaccurate frequency shifts if it relies on an nonphysical or inaccurate glitch modelling. Moreover, if the parameter to be estimated does not appear directly in the model, there is no alternative to calibration on stellar models \citep{Houdek2007,Houdek2011,Verma2014a,Verma2019} which adds its own uncertainty on the internal physics. This method is therefore largely dependent on the work done beforehand and it is in this perspective that this first paper is intended. A second paper will focus on the seismic effects of the structural perturbation and what information these provide about the abundances.\newline

In the present paper, we propose a physical model of the ionisation region that allows us to derive a semi-analytic description of the structural perturbation caused by a change in abundances. In that respect, we will first give a short description of previous modelling in Section \ref{REVIEW}, as well as a more in-depth picture of the most commonly used model of the ionisation glitch, namely \citet{Houdek2007}, hereafter HG07. The formalism of our own model of the ionisation region will be introduced in Section \ref{MOD} and its structural behaviour studied in Section \ref{PARAM}. In Section \ref{DISCUSSION}, we propose an analysis of the structural perturbation induced by the model and how it is related to analytical models used for the ionisation glitch modelling. The last part of the article will be dedicated to our conclusions.

\section{Previous seismic glitch frameworks \label{REVIEW}}

As briefly mentioned, studying seismic glitches requires a twofold approach in the modelling. One must first consider an expression for the structural perturbation and then infer the shape of the associated frequency shift. 

\subsection{Structural perturbation modelling}

\begin{figure}
\centering
\includegraphics[width=\columnwidth]{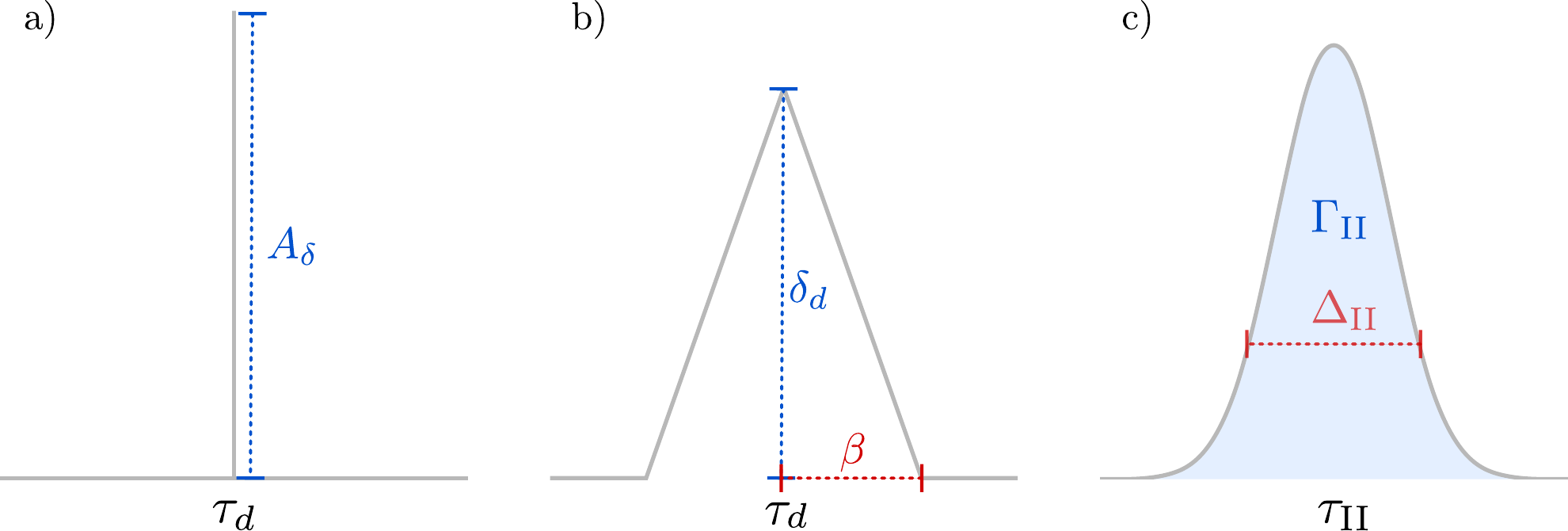}
\caption{Various shapes along with their associated parameters used to describe a structural perturbation. (a) Dirac function used in \cite{Monteiro1994} to model the variations of the acoustic potential (in the case of an overshoot) passing from a convective to a radiative region. For such modelling, it is only necessary to specify the acoustic depth $\tau_d$ and amplitude $A_\delta$ of the perturbation. (b) ``Triangular'' shape of the $\Gamma_1$ perturbation in the second helium ionisation region as in \cite{Monteiro2005}. An additional parameter controls the width ($\beta$) of the perturbation. (c) Gaussian shape of the $\Gamma_1$ perturbation in the second helium ionisation region as in \cite{Houdek2007}. $\Delta_\textrm{II}^2$ and $\Gamma_\textrm{II}$ describe in this case respectively the variance and the area of the distribution.}
    \label{AllGlitches.pdf}%
\end{figure}

The modelling of the perturbation usually boils down to choosing a localised and parametrisable shape for the induced structural change. Thus, depending on the region and the phenomenon under study, various forms can be considered. For the transition between radiative and convective regions, one will generally use very sharp shapes like a Dirac or a step function to reflect a discontinuity in the density derivatives \citep{Monteiro1994,Houdek2007}. However, the ionisation region being more spread out, the shapes used to model the structural perturbations usually involve an additional dispersion parameter \citep{Monteiro2005,Houdek2007}. In Fig. \ref{AllGlitches.pdf}, we represent  shapes introduced in previous papers with their associated parameterisation. In what follows, we will focus on the most commonly used model (i.e. the one presented in HG07) for estimating helium abundances and on the derived expressions for the frequencies.\newline

In HG07, the structural change caused by a change in helium abundance is first modelled  as a Gaussian perturbation of the first adiabatic exponent $\Gamma_1 = (\partial\ln P/\partial\ln \rho)_S$:
\begin{equation}
    \label{eq:dG1_HG07}
    \frac{\delta \Gamma_1}{\Gamma_1} = -\frac{1}{\sqrt{2 \pi}} \frac{\Gamma_{_\textrm{HeII}}}{\Delta_{_\textrm{HeII}}} e ^{-\left(\tau-\tau_{_\textrm{HeII}}\right)^{2} / 2 \Delta_{_\textrm{HeII}}^{2}}
\end{equation} 

\begin{figure}
\centering
\includegraphics[width=8cm]{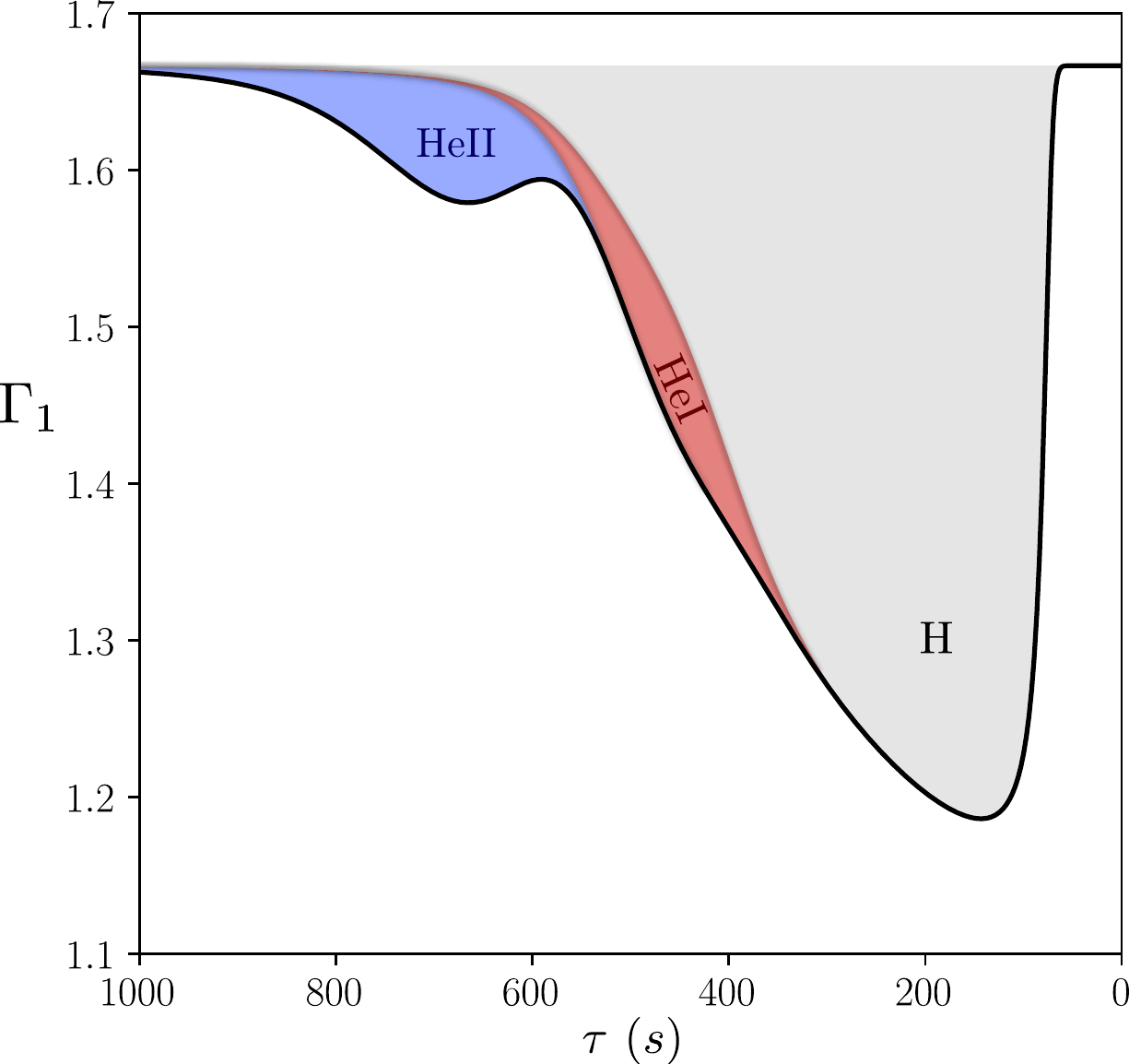}
\caption{Representation of a typical $\Gamma_1$ profile in the ionisation region as a function of the acoustic depth $\tau$ (the surface corresponds to $\tau = 0$). The helium abundance in this plot is $Y \sim 0.25$. The contributions of the three main ionisation zones have been distinguished, i.e. the hydrogen (H), the first (HeI) and second (HeII) helium ionisation zones. Each of these causes a deviation from the $\Gamma_1$ reference value of $5/3$. }
    \label{ThreeWells.pdf}%
\end{figure}

The idea is as follows: the transition from a pure hydrogen model to one partially composed of helium causes the appearance of a well in the $\Gamma_1$ profile (cf. Fig \ref{ThreeWells.pdf}). This results from the second helium ionisation to which the index $\textrm{HeII}$ refers. While various analytical functions can be used to model a well, the latter seems closely reproduced by a Gaussian expressed as a function of the acoustic depth $\tau = \int_r^R dr'/c$; $c(r)$ designating the sound speed at a given position. An important point is that, if the well depth, position or width depends on the helium abundance $Y$ or on the thermodynamic conditions of the CZ, then a connection (though implicit) can be made between physical properties of the CZ and the parameters $\Gamma_{_\textrm{HeII}}$, $\Delta_{_\textrm{HeII}}$ and $\tau_{_\textrm{HeII}}$.\newline

Before further discussion, it may be useful here to clarify an ambiguity concerning the notation ``$\,\delta\,$'' which symbolises the perturbation. Indeed, since it denotes a difference between two \textit{profiles} (say between profiles $\Gamma_1^A$ and $\Gamma_1^B$), the variable over which these are expressed has to be chosen carefully.  For instance, for arbitrary variables $x$ and $y$:
\begin{equation}
    \delta_x \Gamma_1 \equiv \Gamma_1^B(x)-\Gamma_1^A(x) \neq \Gamma_1^B(y)-\Gamma_1^A(y) \equiv \delta_y \Gamma_1
\end{equation}

As shown above, the notation ``$\,\delta_x\,$'' allows us to overcome this ambiguity. The latter has been introduced by \cite{JCD1997} which develop this point extensively. Since the quantity $x$ should vary in the same range for both profiles $A$ and $B$, it is natural to choose it as a normalised variable e.g. $r/R$, $m/M$, $t = \tau/\tau_0$, ... ($R$, $M$, and $\tau_0 = \int_0^R dr/c$ being the total radius, mass, and acoustic radius of the model). Also, one may note that the variable on which the perturbation is expressed can differ from the one that has been used to calculate the difference; both $\delta_x \Gamma_1(x)$ and $\delta_x \Gamma_1(y)$ have a meaning.

This point clarified, it appears that the perturbation used in HG07 for Eq.~\eqref{eq:dG1_HG07} is $\delta_\tau \Gamma_1$ since the Gaussian shape has been established after an expansion at fixed $\tau$ (cf. Eq (31) of HG07). As mentioned, $\delta_\tau$ is however ambiguous if $\tau_0^A \neq \tau_0^B$. It can be replaced by $\delta_t$ following: 
\begin{equation}
    \frac{\delta_t \Gamma_1}{\Gamma_1} = \frac{\delta_\tau \Gamma_1}{\Gamma_1} + \frac{\delta_t\tau}{\tau}\frac{d\ln \Gamma_1}{d\ln\tau} = \frac{\delta_\tau \Gamma_1}{\Gamma_1} + \frac{\delta\tau_0}{\tau_0}\frac{d\ln \Gamma_1}{d\ln\tau}
\end{equation} with $\delta \tau_0 = \tau_0^B - \tau_0^A$ simply being the perturbation of a constant. 

\subsection{Expected frequency shift}

The resulting signature in the frequencies can be divided in two distinct parts:

\paragraph{\emph{Ionisation component}}
The connection between Eq.~\eqref{eq:dG1_HG07} and the frequency shift $\delta\nu_{n,\ell} = \nu_{n,\ell}^B - \nu_{n,\ell}^A$ is made by considering the following asymptotic form \citep{Gough1990}:
\begin{equation}
    \label{eq:kernel_HG07}
    \frac{\delta\nu_{n,\ell}}{\nu_{n,\ell}} = \int_0^1 \left[\mathcal{K}_{\rho c^{2}}^{~n,\ell} \frac{\delta_x \rho}{\rho}+ \mathcal{K}_{c^{2} \rho}^{~n,\ell} \frac{\delta_x c^{2}}{c^{2}}\right] d x \sim \int_0^1 \mathcal{K}_{c^{2} \rho}^{~n,\ell} \frac{\delta_x \Gamma_1}{\Gamma_1} d x
\end{equation} where $\mathcal{K}_{\rho c^{2}}^{~n,\ell}$ and $\mathcal{K}_{c^{2} \rho}^{~n,\ell}$ denote the usual structural kernels that can be found in \cite{Gough1991}. This derivation, as well as the majority of the frequency shift treatment, will be fully discussed in a second paper. Once more, one must pay attention to the perturbation used in \eqref{eq:kernel_HG07}; it can be shown that: 
\begin{equation}
    \frac{\delta_x \Gamma_1}{\Gamma_1} = \frac{\delta_t \Gamma_1}{\Gamma_1} - \left[\frac{\delta \tau_0}{\tau_0}-\frac{\delta R}{R}+\frac{1}{x}\int_{0}^{x} \frac{\delta_{t} c}{c} d x^{\prime}\right]\frac{d\ln\Gamma_1}{d\ln r}
\end{equation} This nuance cannot, however, be appreciated without the notation introduced here. 

The frequency shift derived in HG07 is written as a continuous function of frequency:
\begin{equation}
    \frac{\delta\nu}{\nu} = A_{_\textrm{HeII}}\nu e ^{-8\pi^2 \Delta_{_\textrm{HeII}}^{2} \nu^{2}} \cos 2\left(2\pi\tau_{_\textrm{HeII}} \nu+\epsilon_{_\textrm{HeII}}\right)
\end{equation} as are the second differences:
\begin{equation}
    \label{eq:d2nu_HG07}
    d^2\nu_{_{\Gamma_1}} = F_{_\textrm{HeII}}A_{_\textrm{HeII}}\nu e ^{-8\pi^2 \Delta_{_\textrm{HeII}}^{2} \nu^{2}} \cos \left[2\left(2\pi\tau_{_\textrm{HeII}} \nu+\epsilon_{_\textrm{HeII}}\right)-\delta_{_\textrm{HeII}}\right]
\end{equation} with $A_{_\textrm{HeII}} = \pi\Gamma_{_\textrm{HeII}}/\tau_0$. $F_{_\textrm{HeII}}$ and $\delta_{_\textrm{HeII}}$ are functions of the other parameters and the frequency (though it is assumed that they only fluctuate slowly with it), thus making the total number of parameters only 4.

\paragraph{\emph{Smooth component}}

The helium component is not the only perturbation expected in the second differences. Apart from the signature of the transition between radiative and convective regions, a smooth component $d^2\nu_s$ can also be considered and modelled as: 
\begin{equation}
    \label{eq:smooth}
    d^2\nu_s = \sum_{k=0}^3 a_k \nu^{-k}
\end{equation}

The idea is to first consider a star that would not contain any glitch. Without any structural perturbation (thus simplifying the problem), it can be assumed that its frequencies follow the asymptotic expansion provided by \cite{Tassoul1980}, i.e.:
\begin{equation}
    \label{eq:asym}
    \nu_{n,\ell} = \left[\frac{1}{2}(2n+\ell+\varepsilon)+\frac{2V_\ell}{2n+\ell+\varepsilon}\right]\Delta\nu + \mathcal{O}\left(\frac{1}{{\nu_{n,\ell}}^2}\right)
\end{equation} with $\varepsilon$ and $V_\ell$ two dimensionless parameters independent of the radial order $n$ and $\Delta\nu \equiv 1/2\tau_0$ the asymptotic large frequency separation.
The last term of \eqref{eq:smooth}, which is proportional to $\nu^{-3}$, can then be interpreted as the leading term of the second differences applied to \eqref{eq:asym}:
\begin{equation}
    d^2\nu_{\textrm{asym}} \sim \frac{2V_\ell}{\nu^3} (\Delta\nu)^4
\end{equation}

However, it is difficult to give as convincing a justification for the other terms, which hide a much more complex deviation to the asymptotic expansion. \cite{Houdek2011} list in particular hydrogen ionisation, a sharp stratification of the upper layers of the convection zone, non-adiabatic processes and turbulent perturbations caused by the oscillations. In the end, this component adds 4 free parameters to the ones identified so far. \footnote{It should be noted that the combination of \textit{positive} powers of $\nu$ introduced in \cite{Verma2014a} to describe the smooth component is incompatible with the asymptotic expansion \eqref{eq:asym}.}

\subsection{Improvements and limits of the approach}

Although it is not apparent from this short summary, the derivation of Eqs. \eqref{eq:kernel_HG07}-\eqref{eq:d2nu_HG07} is as complex as intricate, and so an analytical derivation of second differences requires making use of many approximations. The first consequence is that a fit of the second differences with Eq.~\eqref{eq:d2nu_HG07} only allows a very approximate retrieval of the parameters introduced in Eq.~\eqref{eq:dG1_HG07}. HG07 reveal typical discrepancies of $50$, $33$, and $25\%$ on parameters $\Gamma_{_\textrm{HeII}}$, $\Delta_{_\textrm{HeII}}$, and $\tau_{_\textrm{HeII}}$, respectively. These differences can be reduced to $5, 15$ and $15\%$ by introducing a frequency dependence in $\epsilon_{_\textrm{HeII}}$ induced by the cut-off frequency and a more complete model by adding the first helium ionisation contribution in Eq.~\eqref{eq:dG1_HG07} (see Fig. \ref{ThreeWells.pdf}). Having now two Gaussians (indexed by HeI and HeII), the expected second differences become
\begin{equation}
    \label{eq:d2nuv2_HG07}
    d^2\nu_{_{\Gamma_1}} = \sum_{i~ =~\{\textrm{HeI},\,\textrm{HeII}\}} F_iA_i\nu e ^{-8\pi^2 \Delta_i^{2} \nu^{2}} \cos \left[2\left(2\pi\tau_i \nu+\epsilon_i(\nu)\right)-\delta_i\right]
\end{equation} instead of Eq.~\eqref{eq:d2nu_HG07}. Since the parameters of the two Gaussians are not independent, 3 of them are fixed in HG07 by the empirical relations: $A_{_\textrm{HeI}}/A_{_\textrm{HeII}} = 0.5$, $\Delta_{_\textrm{HeI}}/\Delta_{_\textrm{HeII}} = 0.9$ and $\tau_{_\textrm{HeI}}/\tau_{_\textrm{HeII}} = 0.7$. The number of independent parameters in the expression still amounts to 4 (only one is needed to determine both $\epsilon_{_\textrm{HeI}}(\nu)$ and $\epsilon_{_\textrm{HeII}}(\nu)$). However, despite the substantial improvement in the parameter retrieval, \eqref{eq:d2nu_HG07} remains currently the most used equation for studying ionisation glitches \citep{Verma2017,Verma2019,Farnir2019}.\newline

More importantly, the abundances, and in particular the helium abundance $Y$, do not appear directly as parameters in \eqref{eq:d2nuv2_HG07}. As mentioned above, without further theoretical work to establish a dependency between these parameters and $Y$, calibration based on realistic models therefore seems necessary in order to make it appear. This is, in our opinion, the greatest criticism that can be made of the method. To illustrate this point, we will place ourselves in a broader framework and consider two models $\mathcal{P}$ and $\mathcal{M}_\star$ representing respectively the structural perturbation caused by a change in helium $\delta Y$ and the structure of a realistic stellar model. The first is parameterised by a set of values $\Vec{\theta}_p$ and the second by the set $\Vec{\theta}_\star$ from which we will explicitly distinguish $Y$ for its particular role in this context. As an example, if the model $\mathcal{P}$ is chosen to be a Gaussian as done by HG07 (Eq. \eqref{eq:dG1_HG07}), it would then involve three parameters $\Vec{\theta}_p = \left(\Gamma_{_\textrm{HeII}}, \Delta_{_\textrm{HeII}}, \tau_{_\textrm{HeII}}\right)$.  In contrast, $\Vec{\theta}_\star$ will contain instead quantities needed to calculate a realistic model, such as fundamental parameters (mass $M_\star$, radius $R_\star$, age $A_\star$, etc...) but also all the quantities required to model the physical processes involved in the star (overshoot, mixing length, parameters implied in diffusion or rotation, etc... cf. \cite{Lebreton2014} which provides a broad idea of the various possible prescriptions). As a result, $\Vec{\theta}_\star$ generally contains many more components than $\Vec{\theta}_p$. With this in mind, we would like to be able to relate a fitted set of parameters, $\Vec{\theta}_p$, to a difference in helium $\delta Y$ from a chosen reference $Y$. For this purpose, the solution proposed by calibration is to assimilate the \textit{perturbation model} $\mathcal{P}$ to a \textit{model perturbation} $\delta \mathcal{M}$ and therefore to assume:
\begin{equation}
    \label{eq:calibration}
    \mathcal{P}(\Vec{\theta}_p) \simeq \mathcal{M}_\star(Y+\delta Y; \Vec{\theta}_\star) - \mathcal{M}_\star(Y; \Vec{\theta}_\star) \simeq \left.\frac{\partial \mathcal{M}_\star}{\partial Y}\right|_{Y; \Vec{\theta}_\star} \delta Y
\end{equation}
in order to numerically derive a relation $\Vec{\theta}_p(\delta Y)$. It can be noticed that Eq.~\eqref{eq:calibration} is equivalent to Eq. (33) of HG07 when $\mathcal{P}$ is chosen to be the sum of two Gaussians and $Y=0$ as a reference (although the choice of $\Vec{\theta}_\star$ is not specified in HG07). Thus, in order to be able to associate a set of parameters $\Vec{\theta}_p$ to a unique $\delta Y$, the calibration method makes the assumption that $\partial \mathcal{M}_\star/\partial Y$ does not depend on the reference point $(Y; \Vec{\theta}_\star)$. However, despite some arguments that will be given in favour of a relative independence regarding the choice of $Y$, we will show in Section \ref{DISCUSSION} that such an assumption about $\Vec{\theta}_\star$ is inadequate. Such dependencies cannot be reflected in model $\mathcal{P}$ which depends on too few parameters, namely $\Vec{\theta}_p$ (and it should be noted that this is actually part of the reasons for introducing $\mathcal{P}$ in the first place). To minimise the error introduced, it is then necessary to determine as many $\Vec{\theta}_p(\delta Y)$ relations as stars studied (given independent constraints on $\Vec{\theta}_\star$, see \cite{Verma2019}), which can be quite costly. Moreover, by using such a relationship anyway, we end up reintroducing the biases we wanted to avoid when looking at the ionisation glitch. Indeed, the relationship obtained will largely depend on the physics considered in $\mathcal{M}_\star$ and in particular on mechanisms not necessarily relevant for the ionisation region modelling.\newline

To summarise, despite being based on a more sophisticated model than the previous ones, the method described above faces challenges mainly due to its empirical introduction. In particular, the lack of an explicit dependence on $Y$ greatly reduces its applicability. We would like to reverse the approach. The idea is firstly to introduce a physically based parameterisation allowing inferences concerning ionisation regions and secondly to give a mathematical description regarding the $\Gamma_1$ profile. The challenge will be to provide such a modelling while keeping the number of parameters as low as possible. This whole issue will be addressed in the two following sections.

\section{First adiabatic exponent in an ionisation region \label{MOD}}

In order to model the structure of the ionisation zone, we will first try to provide an analytical expression of $\Gamma_1$ in this specific region. As previously mentioned, we will avoid introducing it via similarities with mathematical functions but rather derive it from thermodynamic relationships. Such expressions have already been obtained \citep[e.g.][]{Cox1968,Kippenhahn2012}, but are generally functions of the occupation number rather than state variables (such as $T$ and $V$). Furthermore, due to the complexity of the chemical equilibrium resulting from Saha's equations, these equations are generally solved numerically, and analytical expressions are limited to the ionisation of single species. We will try to present in the following the simplest model that can nevertheless explicitly involve chemical composition.

\subsection{Free energy}

Our starting point is the free energy of a multi-component ideal gas, i.e.:
\begin{equation}
    \label{eq:Fid}
    F(T,V,N_\alpha) = \sum_\alpha N_\alpha (\mu_\alpha -kT)
\end{equation} where $V$ designates the volume, $kT$ the temperature expressed in energy units, and $N_\alpha$ and $\mu_\alpha(T,V,N_\alpha)$ the number and chemical potential of particles of type $\alpha$. In the context of partial ionisation, $\alpha$ refers to $(i,r)$, $1 \leq i \leq \spec$ being an index on the chemical species and $0 \leq r \leq z_i$ the ionisation state (where $z_i$ is the atomic number), or may also correspond to $e$ for an electron. The key point of Eq.~\eqref{eq:Fid} lies in its ability to provide us with the pressure $P$ and entropy $S$ via its first derivative. Moreover, its second derivative gives access to virtually any other thermodynamic quantity ranging from the speed of sound, through the adiabatic exponents including $\Gamma_1$, to various compressibilities. Before diving into the calculations, we would like to take a closer look at the particular meaning of a derivative in stellar physics compared with pure thermodynamics. The first adiabatic exponent is often defined as follows\footnote{The derivation of $\Gamma_1$ will be performed using the more convenient state variable $V$ instead of $\rho$.}:
\begin{equation}
    \label{eq:G1def1}
    \Gamma_1 \equiv \left(\frac{\partial \ln P}{\partial \ln \rho}\right)_S = - \left(\frac{\partial \ln P}{\partial \ln V}\right)_S
\end{equation}where the subscript $S$ denotes a partial derivative taken at constant entropy. The above expression can easily be rewritten using the more common variables $T$ and $V$:
\begin{equation}
    \label{eq:G1def2}
    \begin{split}
        \Gamma_1 
        = \frac{V}{P} {\left(\frac{\partial S}{\partial T}\right)_{V}}^{-1}\left[\left(\frac{\partial P}{\partial T}\right)_{V}\left(\frac{\partial S}{\partial V}\right)_{T}-\left(\frac{\partial P}{\partial V}\right)_{T}\left(\frac{\partial S}{\partial T}\right)_{V}\right]
    \end{split}
\end{equation} 

Although this equation is generally valid, its use in the present case will require some clarification as it only mentions the two state variables $T$ and $V$ and does not explicitly show the dependence on the various number of particles $N_\alpha$. Assuming that species are indexed so that $z_i = i$ ($N_i = \sum_r N_i^r$ potentially being null), the free energy introduced in Eq.~\eqref{eq:Fid} is a function of $\spec(\spec+3)/2+3$ variables. As a result, the partial derivatives of Eqs. \eqref{eq:G1def1} \& \eqref{eq:G1def2} are generally ambiguous unless some indication is given concerning the $\spec(\spec+3)/2+1$ implicit conservation equations. It is usual to consider partial derivatives at constant values of state variables, e.g.: 
\begin{equation}
    \label{eq:d1F}
    S = -\left(\frac{\partial F}{\partial T}\right)_{V,N_\alpha}~,\quad
    P = -\left(\frac{\partial F}{\partial V}\right)_{T,N_\alpha}~,\quad
    \forall \alpha\quad \mu_\alpha = \left(\frac{\partial F}{\partial N_\alpha}\right)_{T,V,N_{\beta\neq\alpha}}
\end{equation}

However, it is clear that the derivatives appearing in Eq.~\eqref{eq:G1def2} (and second derivatives of $F$ considering Eq.~\eqref{eq:d1F}) are not of this kind. As mentioned in the introductory section, the $\Gamma_1$ profile used in stellar physics shows telltale signs of ionisation that derivatives taken at constant $N_\alpha$ do not. Indeed, this choice imposes a particular ionisation state in the whole region (described by the constant $N_\alpha$ values), which instead should fluctuate widely with the thermodynamic conditions in the CZ. The solution is to instead consider the following conservation equations:
\begin{align}
    \label{eq:Eq1}
    \forall i,~\forall r>0, \quad& \mu_i^r+\mu_e-\mu_i^{r-1} = 0 \qquad\qquad \left[~\frac{\spec(\spec+1)}{2}~\right] \\
    \label{eq:Eq2}
    \forall i, \quad& \sum_{r=0}^{z_i} N_i^r = N_i = \textrm{cnst} \qquad\qquad\quad~~~ \left[~\spec~\right]\\
    \label{eq:Eq3}
    \quad& \sum_{i=1}^{\spec}\sum_{r=1}^{z_i} rN_i^r = N_e \qquad\qquad\qquad\quad \left[~1~\right]
\end{align} The first one corresponds to the chemical equilibrium of the ionisation reaction: $A_i^{r-1} \rightleftharpoons A_i^r + e^-$.  The second is merely the conservation of each atom number and can also be considered as a chemical equilibrium in the absence of reactions and homogeneous mixing of the CZ. The last expression corresponds to the overall charge equilibrium, i.e. electroneutrality. This set of relations can be interpreted as a local equilibrium for given temperature and volume conditions that should be verified at each point of the CZ, and partial derivatives taken with respect to these constraints will hereafter be denoted with the subscript ``EQ''. The resulting number of constraints has been indicated in square brackets and add up to $\spec(\spec+3)/2+1$ as required. 

Hereafter, we will use the notation $\partial_{\alpha\beta}^2 F$ to designate a second derivative taken with respect to $\beta$ (subject to the $N_{\alpha}$ being constant) and then $\alpha$ (subject to EQ). For example, Eq.~\eqref{eq:G1def2} becomes:
\begin{equation}
    \label{eq:G1d2F}
    \Gamma_1 \equiv - \left(\frac{\partial \ln P}{\partial \ln V}\right)_{S,EQ} = V~\frac{(\partial_{VT}^2F)(\partial_{TV}^2F)-(\partial_{VV}^2F)(\partial_{TT}^2F)}{(\partial_{V}F)(\partial_{TT}^2F)}
\end{equation}by using relations \eqref{eq:d1F}. Considering Eq.~\eqref{eq:Fid}, the evaluation of the Eq.~\eqref{eq:G1d2F} now boils down to finding the $N_i^r(T,V)$ at a given EQ.

\subsection{Approximate local equilibrium \label{EQ}}

To solve the system \eqref{eq:Eq1},\eqref{eq:Eq2}, \eqref{eq:Eq3}, one must first explicitly write out the expression of the chemical potential of each particle. We have:
\begin{equation}
    \label{eq:mualpha}
    \mu_\alpha(T,V,N_\alpha) = kT\ln\left(\frac{N_\alpha\lambda_\alpha^3}{\mathcal{Z}_\alpha V}\right)
\end{equation}with $\lambda_\alpha(T)$ and $\mathcal{Z}_\alpha(T)$ the thermal De Broglie wavelength and the canonical partition function of a particle $\alpha$, respectively. Conditions $\eqref{eq:Eq1}$ thus become:
\begin{equation}
    \label{eq:Saha1}
    \forall i, \forall r>0,\quad \frac{x_i^r}{x_i^{r-1}} = \frac{\mathcal{Z}_i^r\mathcal{Z}_e}{\mathcal{Z}_i^{r-1}}\frac{V}{N_e{\lambda_e}^3}
\end{equation} with $x_i^r = N_i^r/N_i$ the occupation number of state ($i$,$r$). In the ideal case\footnote{In our effort to obtain the simplest model that can be used to handle chemical composition, the ideal gas assumption remains relevant. It should be noted, however, that in order to take into account the ionisation process in a more accurate way, it would be necessary to consider the Coulombian effects \citep{Rogers1981,Rogers1996}.}, $\mathcal{Z}_i^r$ and $\mathcal{Z}_e$ are often approximated by \citep[e.g.][]{Kippenhahn2012}:
\begin{align}
    \mathcal{Z}_i^r(T) &\simeq u_i^r(T)\exp\left(\displaystyle \sum_{s=1}^r\chi_i^s/kT\right) \simeq g_i^r\exp\left(\displaystyle \sum_{s=1}^r\chi_i^s/kT\right)\\
    \label{eq:Ze}
    \mathcal{Z}_e &\simeq 2
\end{align}with $u_i^r$ describing the fine structure of state ($i$,$r$) approached by the degeneracy of the ground state $g_i^r$ and $\chi_i^r$ the ionisation energy separating state ($i$,$r$) from ($i$,$r-1$) ($\sum_{s}^r\chi_i^s/kT$ is then the energy from the reference state ($i$,$0$) assumed here to have a null internal energy). The second approximation \eqref{eq:Ze} directly results from neglecting the electron spin energy contribution.

Thus, conditions \eqref{eq:Saha1} simply become Saha's equations:
\begin{equation}
    \label{eq:Saha2}
    \forall i, \forall r>0,\quad \frac{x_i^r}{x_i^{r-1}} = \frac{2g_i^r}{g_i^{r-1}}\frac{V}{N_e{\lambda_e}^3}e^{-\chi_i^r/kT}
\end{equation}

These equations are highly coupled, especially because of the term $N_e$ that can be expressed from \eqref{eq:Eq3} as:
\begin{equation}
    N_e = N \sum_{i=1}^{\spec}\sum_{r=1}^{z_i} rx_ix_i^r 
\end{equation} with $N = \sum_i N_i$ and $x_i = N_i/N$ respectively the total number of atoms and the \textit{number} abundance of element $i$, which are two constants from \eqref{eq:Eq2}. Because of this, Saha's equations are generally solved numerically by iterating over estimates of the $x_i^r$. In the following, we will try to give analytical approximations by considering a few simplifying assumptions. First, let us introduce the cumulative occupation number $y_i^r = \sum_{s\geq r} x_i^s$, more suited to the study of Eqs. \eqref{eq:Saha2} \citep{Baker1962}. Also, we will use from this point the convention: $$\sum_{ir} \equiv \sum_{i=1}^{\spec}\sum_{r=1}^{z_i}$$

This way, one can write $N_e = N\Bar{z}$ with:
\begin{equation}
    \label{eq:zbar}
    \Bar{z} = \sum_{ir}x_iy_i^r
\end{equation} the mean number of electrons per atom. Saha's equations can be rewritten:
\begin{equation}
    \label{eq:Saha3}
    \forall i, \forall r>0,\quad \frac{y_i^r-y_i^{r+1}}{y_i^{r-1}-y_i^r} = \frac{2g_i^r}{g_i^{r-1}}\frac{V}{N\Bar{z}{\lambda_e}^3}e^{-\chi_i^r/kT}
\end{equation}with the convention $y_i^{z_i+1} = 0$ (and by definition $y_i^0 = 1$). Let us start by noting that it is rare to find more than 2 ionisation states simultaneously for a given element. It is thus advantageous to make the following approximation \citep{Baker1962}:
\begin{equation}
    \begin{split}
        \forall i, \forall r>0,~\textrm{if}~y_i^r \neq 0,~\textrm{then}, \quad \forall s<r, y_i^s&=1 \\
        \forall s>r, y_i^s&=0
    \end{split}
\end{equation}

   \begin{figure}
   \centering
   \includegraphics[width=9cm]{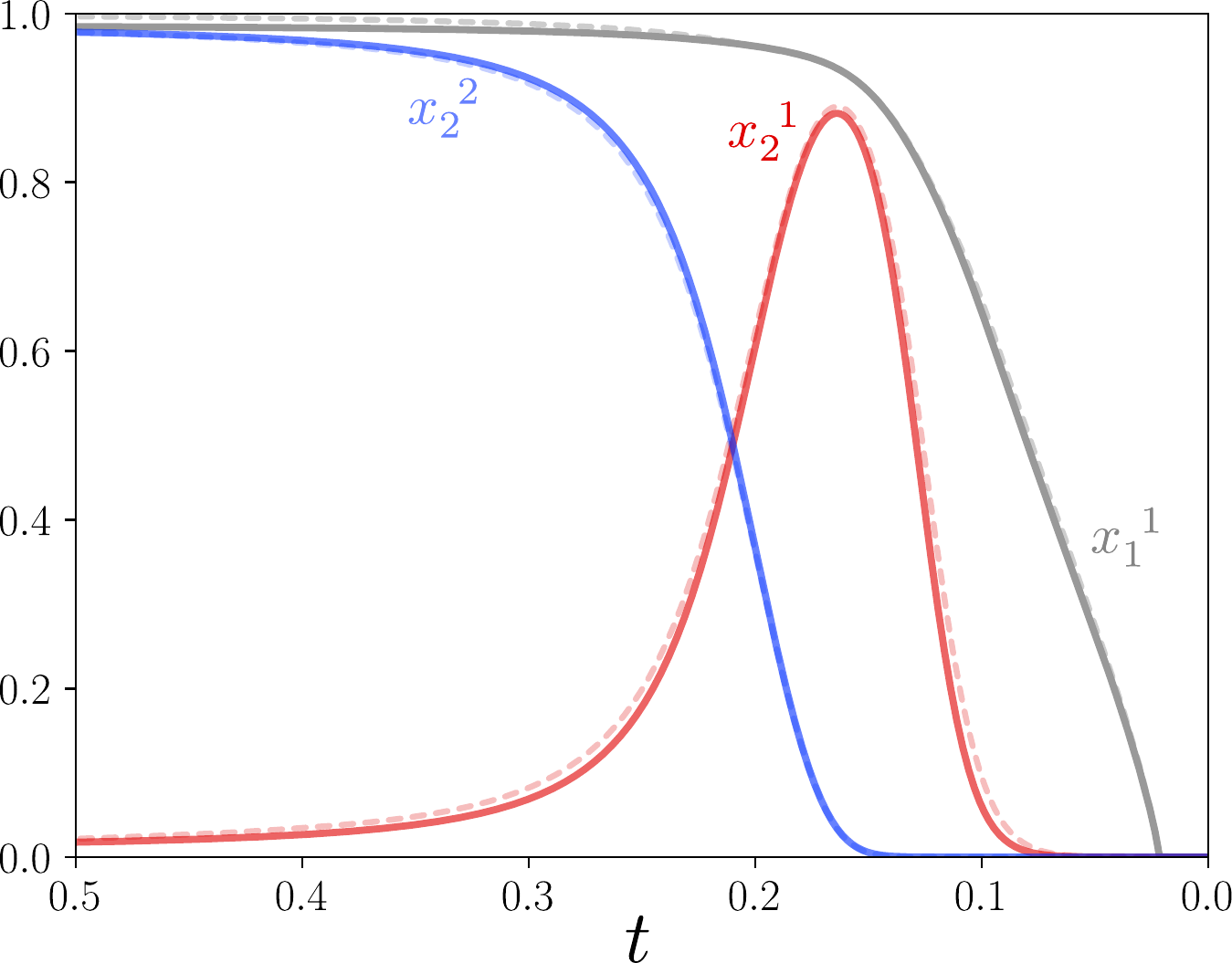}
      \caption{Comparison of approximations \eqref{eq:y11} and \eqref{eq:yir} in the case $i=1,2$ (opaque curves) with numerical solution of \eqref{eq:Saha2} (dashed light curves) for solar near-surface conditions of ($T,V,X_{i>1}$). Note that these quantities are represented as a function of the normalised acoustic depth $t = \tau/\tau_0$ (making the surface located at $t=0$), along which all future plots will be represented.}
         \label{Allxir.pdf}
   \end{figure}

Then, in the single domain of interest ($y_i^r \neq 0$), \eqref{eq:Saha3} will be approached by:
\begin{equation}
    \label{eq:Sah4}
    \forall i, \forall r>0,\quad \frac{y_i^r}{1-y_i^r} = \frac{2g_i^r}{g_i^{r-1}}\frac{V}{N\Bar{z}{\lambda_e}^3}e^{-\chi_i^r/kT}
\end{equation}

A final assumption is needed to express $\Bar{z}$ as a function of $y_i^r$ only and to completely decouple the system. One idea may be to exploit the fact that stars are predominantly composed of hydrogen in number: for typical $Y$ values (i.e. $1/4<Y<1/3$), $x_1\sim 0.89-0.92$ and it can be seen using \eqref{eq:zbar} that after a complete ionisation of the hydrogen $\Bar{z}>x_1$. Another bound can then be obtained on $\bar{z}$ since the latter tends asymptotically towards $\sum_i z_ix_i = 2-x_1+\sum_{i>2}(z_i-2)x_i$. Because for heavy elements ($i>2$), $m_i \simeq 2z_im_u$ with $m_i$ the mass of atom $i$ and $m_u$ the atomic mass unit, and since the mean mass $m_0 < 2m_u$ in a stellar mixture, we have $m_0/m_i < 1/z_i$. Finally, the number and mass abundances are related by $x_i = m_0 X_i/m_i$ meaning that $\bar{z}$ should remain below the upper boundary $2-x_1+Z \sim 1.1$ with $Z=\sum_{i>2}X_i$ the metal abundance in mass. This simple reasoning allows us to convince ourselves that $\bar{z}$ must remain close to $1$ once the ionisation of hydrogen completed, and in a way relatively independent of the mixture under consideration.

A pragmatic assumption that can therefore be made is that the ionisation of hydrogen happens first and finishes before any other ionisation begins. Thus we can use the fact that hydrogen is the dominant component to approximate $\bar{z}$ by the mean electron number in a full hydrogen model. Despite not being entirely true (and particularly for a low state of ionisation), this assumption greatly simplifies the problem. Indeed, for all elements $i>1$, $\bar{z}\sim 1$ which is, as discussed above, a reasonable approximation. For the ionisation equation corresponding to hydrogen ($i=1$), we will instead approximate $\bar{z} \sim y_1^1$. Our version of Saha's equations can thus finally be written as:
\begin{equation}
    \label{eq:Saha5}
    \forall i, \forall r>0,\quad \frac{{(y_i^r)}^{n}}{1-y_i^r} = \frac{2g_i^r}{g_i^{r-1}}\frac{V}{N{\lambda_e}^3}e^{-\chi_i^r/kT}
\end{equation}with $n=2$ if $i=1$ and $n=1$ in any other case. Let us define $K_i^r(T,V)$ as the term on the right hand side (RHS) of \eqref{eq:Saha5}. We then obtain the desired relationships:
\begin{align}
    \label{eq:y11}
    y_1^1(T,V) &= \frac{1}{2}\left[\sqrt{K_1^1(K_1^1+4)}-K_1^1\right]\\
    \label{eq:yir}
    \forall i>1, \forall r>0,\quad y_i^r(T,V) &= \frac{K_i^r}{1+K_i^r}
\end{align}

One can easily retrieve the $x_i^r(V,T)$ from there using $x_i^r~=~y_i^r-y_i^{r+1}$ (as well as the $N_i^r(V,T)$ using $N_i^r = N_ix_i^r$). Fig. \ref{Allxir.pdf} shows a comparison of hydrogen and helium ionisation fractions $x_1^1, x_2^1$, and $x_2^2$ obtained analytically from \eqref{eq:y11} and \eqref{eq:yir} for given $T$ and $V$ values with the numerical solution of Saha's Eqs. \eqref{eq:Saha2}. The result is shown as a function of the normalised acoustic depth $t$ so that it can be reduced to a comparison of profiles rather than a $(V,T)$ mapping. The solution presented above seems satisfying considering the number of assumptions made. One may note, however, that $x_2^1$ is slightly underestimated because $\bar{z}$ is actually a little lower than $1$ in the first helium ionisation region. The opposite reasoning holds for the second helium ionisation region.

\subsection{First adiabatic exponent}

Approximations \eqref{eq:y11} and \eqref{eq:yir} allows us to derive all derivatives appearing in \eqref{eq:G1d2F}. First of all, we have:
\begin{equation}
    \label{eq:dVF}
    \partial_VF = -P = \frac{-kT}{V}\sum_\alpha N_\alpha = \frac{-NkT}{V}\left(1+\sum_{ir}x_iy_i^r\right)
\end{equation}where we explicitly made the $y_i^r$ apparent to fully exploit relationships \eqref{eq:y11} and \eqref{eq:yir}. It is straightforward to see that we simply retrieve the classical expression for the pressure in an ideal gas. Second derivatives taken at EQ ensue:
\begin{align}
    \partial_{VV}^2 F &= \frac{NkT}{V^2}\left[1+\sum_{ir}x_iy_i^r\left(1-\frac{\partial\ln y_i^r}{\partial\ln V}\right)\right]\\
    \partial_{TV}^2 F &= \frac{-Nk}{V}\left[1+\sum_{ir}x_iy_i^r\left(1+\frac{\partial\ln y_i^r}{\partial\ln T}\right)\right]
\end{align}

More explicit expressions can be deduced by writing $\displaystyle \frac{\partial\ln y_i^r}{\partial\ln \alpha} = \frac{d\ln y_i^r}{d\ln K_i^r}\frac{\partial\ln K_i^r}{\partial\ln \alpha}$ and exploiting Eq. \eqref{eq:Saha5}
\begin{align}
    \forall i, \forall r>0,\quad \frac{d\ln y_i^r}{d\ln K_i^r} &= \frac{1-y_i^r}{1+\delta_i^1(1-y_i^r)}\\
    \frac{\partial\ln K_i^r}{\partial\ln V} &= 1\\
    \frac{\partial\ln K_i^r}{\partial\ln T} &=\frac{3}{2}+\frac{\chi_i^r}{kT} \equiv \phi_i^r
\end{align}$\delta_i^j$ standing for Kronecker's symbol. Resulting expressions of the derivatives are:
\begin{align}
    \partial_{VV}^2 F &= \frac{NkT}{V^2}\left[1+\sum_{ir}x_iy_i^r\left(1-\frac{1-y_i^r}{1+\delta_i^1(1-y_i^r)}\right)\right]\\
    \partial_{TV}^2 F &= \frac{-Nk}{V}\left[1+\sum_{ir}x_iy_i^r\left(1+\frac{(1-y_i^r)\phi_i^r}{1+\delta_i^1(1-y_i^r)}\right)\right]
\end{align}

The calculation of the two remaining derivatives is however more subtle. We propose here to start with a determination of the energy $E$:
\begin{equation}
    \label{eq:E}
    E = -T^2\left(\frac{\partial}{\partial T}\left(\frac{F}{T}\right)\right)_{V,N_\alpha} = NkT\left(\frac{3}{2}+\sum_{ir}x_iy_i^r\phi_i^r\right)
\end{equation}

Since $\partial_{VT}^2F = -(\partial S/\partial V)_{T,EQ}$ and $\partial_{TT}^2F = -(\partial S/\partial T)_{V,EQ}$, one might be tempted to take directly the derivatives of \eqref{eq:E} using $dE=TdS-PdV$. However, in the present framework:
\begin{equation}
    \label{eq:dE}
    \begin{split}
       dE &= TdS-PdV+\sum_\alpha\mu_\alpha dN_\alpha \\
       &= TdS-PdV+N\sum_{ir}x_i(\mu_i^r+r\mu_e)dx_i^r+N\sum_ix_i\mu_i^0dx_i^0\\
       &= TdS-PdV+N\sum_{ir}x_i(\mu_i^r+\mu_e-\mu_i^{r-1})dy_i^r\\
       &~~~+N\sum_{ir}x_i\mu_i^0dx_i^r+N\sum_ix_i\mu_i^0dx_i^0
    \end{split}
\end{equation}

Clearly, the last two sums cancel each others since $\forall i,~ \sum_{r>0}dx_i^r = d(1-x_i^0) = -dx_i^0$, which is verified in our derivation as long as we define $x_i^0 \equiv 1-y_i^1$. Electroneutrality, used in line 2 of Eq.~\eqref{eq:dE}, is also verified by imposing $N_e \equiv N\sum_{ir}x_iy_i^r$. According to Eq.~\eqref{eq:Eq1}, $\mu_i^r+\mu_e-\mu_i^{r-1} = 0$ and the third term must also cancel exactly. In practice, Eq.~\eqref{eq:Eq1} is too complex to be perfectly solved and numerical calculations as well as analytical approximations can lead to some small departures from exact cancellation. These may lead to slight thermodynamic inconsistencies (which could be corrected thanks to departures from equality of partial mixed second derivatives of state functions like the free energy $F$). However these residuals remain small and we will consider here that chemical equilibrium is perfectly satisfied. Therefore, the only part that remains from \eqref{eq:dE} is the following well known identity:
\begin{equation}
    dE = TdS-PdV
\end{equation}

This means that $T\partial_{VT}^2F = -(\partial E/\partial V)_{T,EQ}-P$ and $T\partial_{TT}^2F~=~-(\partial E/\partial T)_{V,EQ}$. It follows from Eq.~\eqref{eq:E}:
\begin{align}
    \partial_{VT}^2 F &= \frac{-Nk}{V}\left[1+\sum_{ir}x_iy_i^r\left(1+\frac{(1-y_i^r)\phi_i^r}{1+\delta_i^1(1-y_i^r)}\right)\right]\\
    \partial_{TT}^2 F &= \frac{-Nk}{T}\left[\frac{3}{2}+\sum_{ir}x_iy_i^r\left(\frac{3}{2}+\frac{(1-y_i^r){(\phi_i^r)}^{2}}{1+\delta_i^1(1-y_i^r)}\right)\right]
\end{align} 

Denoting the bracket part of $\partial^2_{\alpha\beta}F$ as  $\partial^2_{\alpha\beta}f$, one obtains:
\begin{equation}
    \label{eq:G1d2f}
    \Gamma_1 = \frac{(\partial_{VT}^2f)(\partial_{TV}^2f)+(\partial_{VV}^2f)(\partial_{TT}^2f)}{(\partial_{V}f)(\partial_{TT}^2f)},
\end{equation}where
\begin{align}
    \partial_Vf &= 1+\sum_{ir}x_iy_i^r,\\
    \label{eq:dVVf}
    \partial_{VV}^2 f &= 1+\sum_{ir}x_iy_i^r\left(1-\frac{1-y_i^r}{1+\delta_i^1(1-y_i^r)}\right),\\
    \partial_{TV}^2 f = \partial_{VT}^2 f &= 1+\sum_{ir}x_iy_i^r\left(1+\frac{(1-y_i^r)\phi_i^r}{1+\delta_i^1(1-y_i^r)}\right),\\
    \label{eq:dTTf}
    \partial_{TT}^2 f &= \frac{3}{2}+\sum_{ir}x_iy_i^r\left(\frac{3}{2}+\frac{(1-y_i^r){(\phi_i^r)}^{2}}{1+\delta_i^1(1-y_i^r)}\right).
\end{align}

The various symmetries present in these equations may give the impression that expression \eqref{eq:G1d2f} can be further simplified. In fact, we show in appendix \ref{A} that $\Gamma_1$ can finally be written as:
\begin{equation}
    \label{eq:G1g1}
    \Gamma_1 = \frac{5}{3} - \frac{2}{3}\gamma_1
\end{equation}with $0\leq\gamma_1<1$ simply expressed as:
\begin{equation}
    \label{eq:g1}
    \gamma_1 = \frac{1}{\partial_{TT}^2f}\sum_{ir}x_iy_i^r(1-y_i^r)\frac{{(\chi_i^r/kT)}^{2}}{1+\delta_i^1(1-y_i^r)}
\end{equation}

\begin{figure*}
\centering
\includegraphics[width=\textwidth]{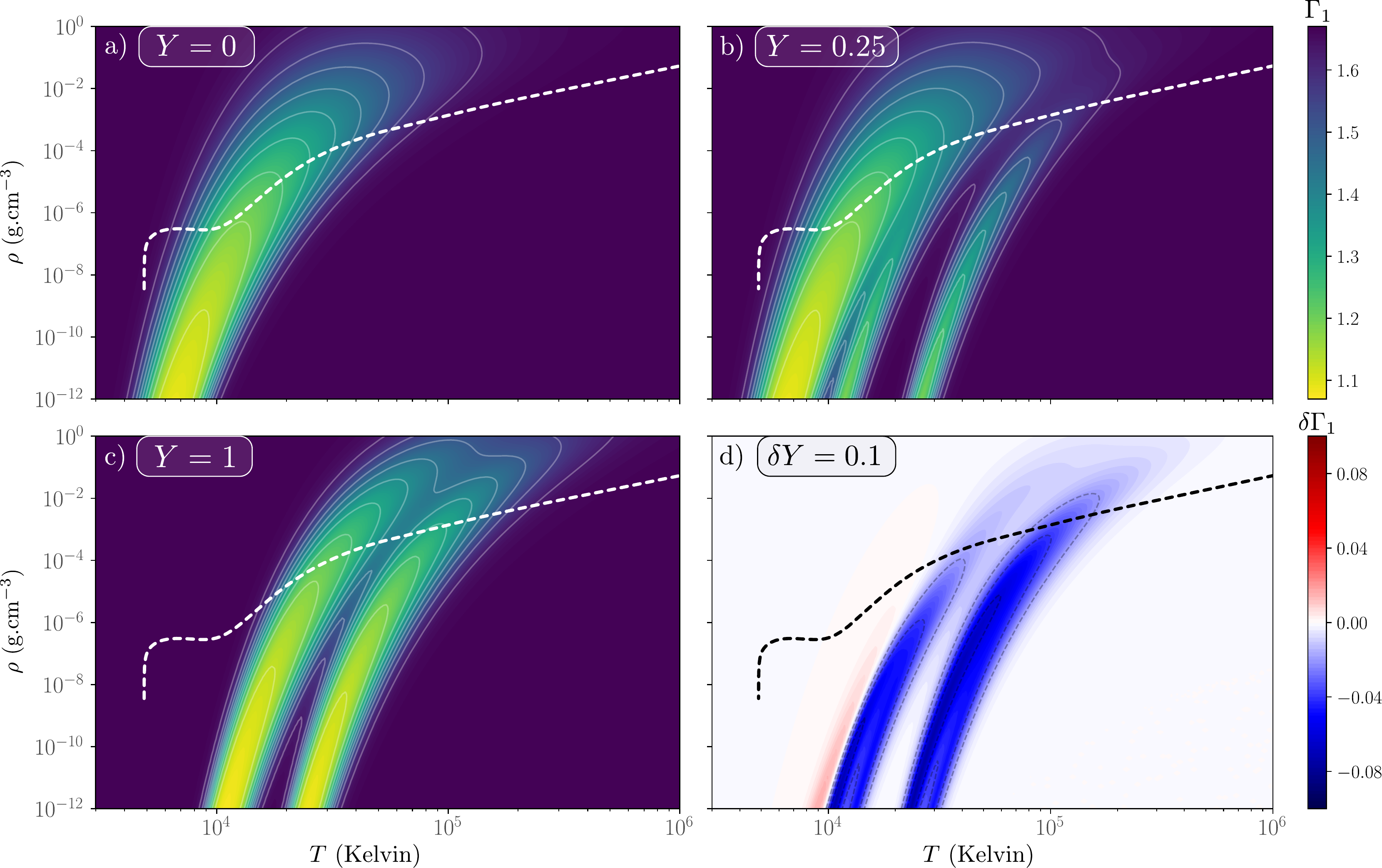}
\caption{Panels (a), (b) and (c): $\Gamma_1(\rho,T,Y)$ as a function of $\rho$ and $T$ for different values of $Y$. Panel (d) represents the variation $\delta_{\rho,T}\Gamma_1$ caused by a perturbation $\delta Y = 0.1$ from a reference value $Y=0.25$. The 
dashed line present on each panel shows the relation $\rho(T)$ extracted from a CESTAM solar model.}
    \label{MapG1.pdf}%
\end{figure*}

Together with Eqs.~\eqref{eq:Saha5}-\eqref{eq:yir}, Eqs. \eqref{eq:dTTf}-\eqref{eq:g1} provide an analytical approximation of $\Gamma_1(T,V,x_i)$ in the CZ. Even if this expression is applicable to any kind of mixture (though assumed to be homogeneous in the CZ!), the following examples will be based on a hydrogen-helium mixture of mass fraction $1-Y$ and $Y$ ($Y = 4x_2/(1+3x_2)$) to facilitate the study. Remembering that $N/V = \rho/m_0$, we gave a representation of the resulting $\Gamma_1(\rho,T,Y)$ in Fig. \ref{MapG1.pdf} for $Y=0.25$ as well as for extreme values, $Y=0$ and $Y=1$. Each panel also shows the relation $\rho(T)$ extracted from a CESTAM solar model \citep{Morel2008,Marques2013}. The latter is obviously only indicative as it depends on the composition, but nevertheless provides an idea of what part of the map is visible on a $\Gamma_1$ profile. The figure clearly reflects the $\Gamma_1$ depressions caused by (from left to right) the hydrogen, and first and second helium ionisations.  Each element contribution is enhanced in the left panels. In the top-right panel, one can notice the superposition of the hydrogen and helium first ionisation zones in typical solar conditions whereas the distinction becomes apparent at lower densities. In contrast, this signature seems to be more diffuse at the highest densities. This behaviour appears in each panel and will be further discussed in Section \ref{PARAM}. The last frame shows the variation $\delta_{\rho,T}\Gamma_1$ caused by a change of helium abundance, $\delta Y = 0.1$, from a reference value $Y=0.25$. One may note that, even if the variables are not normalised, the notation ``$\delta_{\rho,T}$'' is still meaningful since $\Gamma_1$ does not refer here to a profile but a function defined for all $(\rho,T)$ at a given $Y$. As expected, such a variation in abundance results in a lowering of the two helium wells while elevating the hydrogen one. However, the variations seem fairly disproportionate between the two elements. In fact, the change in the hydrogen ionisation region would be barely noticeable considering the typical solar relation $\rho(T)$ shown in black. Appendix \ref{A} provides clues that help to understand this difference in behaviour as well as the particular shape of the perturbation in the hydrogen region compared with the helium one. In particular, the amplitude of the variation seems more related to the relative change $\delta x_i / x_i$ than $\delta Y$. While we have $\delta x_1 = -\delta x_2$ for the hydrogen-helium case, we see that $\delta x_1 / x_1 = -\delta x_2/x_1 = - (x_2/x_1) \delta x_2 / x_2$ with $(x_2/x_1) = 1/12$ for $Y=0.25$. This explains why the change in hydrogen is expected to be an order of magnitude lower than the helium one.\newline

Although Eqs.~\eqref{eq:G1g1} and \eqref{eq:g1} seem promising, they are merely functional forms and do not allow us at this point to model a parameterised structure of the ionisation region. However, $\Gamma_1$, $\rho$ and $T$ are closely related in stellar interiors.

\section{Model structure and properties\label{PARAM}}

In this section we will obtain the stellar structure associated with the $\Gamma_1$ expression derived in the previous section (see Eqs.~\eqref{eq:G1g1} and~\eqref{eq:g1}). For that, we place ourselves in the conditions of a CZ, which correspond to the assumptions while deriving the first adiabatic exponent:
\begin{enumerate}
    \item Isentropic region
    \item Chemical equilibrium at any position
    \item Uniform abundances (due to the homogeneous mixing)
    \item Electroneutrality at any position
\end{enumerate}The last 3 correspond to applying EQ conditions at any point in the region. Also, to avoid getting lost in the many possibilities offered by these relations in terms of the mixture, we will once more consider a relation of the form $\Gamma_1(\rho,T,Y)$ in this part. It is however possible to generalise all the observations made here to any (reasonable) mixture by replacing $Y$ by $X_{i>1}$.

\subsection{Ionisation region structure}

As mentioned above, the quantities $\Gamma_1$, $\rho$ and $T$ are not independent in stellar interiors and, although it may be interesting to study $\Gamma_1$ at any condition of temperature and density, relevant profiles in these environments are actually much more constrained. The appropriate equations to express these constraints are the well-known hydrostatic equilibrium and Poisson's equation assuming spherical symmetry:
\begin{align}
    \label{eq:hydrostatic}
    \frac{dP}{dr} &= -\rho g\\
    \label{eq:Poisson}
    \frac{dg}{dr}+\frac{2g}{r} &= 4\pi G\rho
\end{align} where we have introduced the additional intermediate variable $g$, the norm of the gravity field. At this point, it is important to keep in mind here that we are trying to account for the variation of $T$ and $\rho$ in the region rather than that of $P$. This can be achieved by noticing that, for any quantity $\alpha$, $\displaystyle \frac{d\alpha}{dr} = \frac{-\rho g \alpha}{P}~{\left(\frac{d\ln P}{d\ln \alpha}\right)}^{-1}$. Then, crucially, we note that derivatives taken with respect to the radius $r$ such as $d \ln P/d\ln \alpha = (d\ln P/dr) \,(d\ln \alpha/dr)^{-1}$ exactly correspond to derivatives taken at constant entropy and EQ (cf. assumptions 1-4 at the beginning of the section). Using this for $\alpha = T, \rho$, Eqs. \eqref{eq:hydrostatic} and \eqref{eq:Poisson} can be changed into a differential system of 3 equations, the solution of which is $(T,\rho,g)$:
\begin{align}
    \frac{dT}{dr} &= -\frac{\rho T g}{P}\frac{\Gamma_2 - 1}{\Gamma_2}\\
    \label{eq:drhodr}
    \frac{d\rho}{dr} &= -\frac{\rho^2 g}{P}\frac{1}{\Gamma_1}\\
    \frac{dg}{dr} &= 4\pi G\rho - \frac{2g}{r}
\end{align} where we introduced the second adiabatic exponent $\Gamma_2$ such that $\displaystyle \frac{\Gamma_2}{\Gamma_2-1} \equiv \left(\frac{\partial\ln P}{\partial\ln T}\right)_{S,EQ}$. Its expression can easily be deduced from $\Gamma_1$ by noticing that their definitions are symmetrical with respect to a change $V \leftrightarrow T$:
\begin{equation}
    \begin{split}
        \frac{\Gamma_2}{\Gamma_2-1} &= \frac{T}{P} {\left(\frac{\partial S}{\partial V}\right)_{T}}^{-1}\left[\left(\frac{\partial P}{\partial T}\right)_{V}\left(\frac{\partial S}{\partial V}\right)_{T}-\left(\frac{\partial P}{\partial V}\right)_{T}\left(\frac{\partial S}{\partial T}\right)_{V}\right] 
    \end{split}
\end{equation}

Note that an analogous expression to Eq. \eqref{eq:G1g1} can be provided for $\Gamma_2$:
\begin{equation}
    \frac{\Gamma_2}{\Gamma_2-1} = \frac{5}{2} + \gamma_2
\end{equation}by introducing 
\begin{equation}
    \gamma_2 = \frac{1}{\partial_{VT}^2f}\sum_{ir}x_iy_i^r(1-y_i^r)\frac{(\chi_i^r/kT)(5/2+\chi_i^r/kT)}{1+\delta_i^1(1-y_i^r)}
\end{equation}

One can also get rid of the pressure using the ideal gas law \eqref{eq:dVF} which leads to the following system:
\begin{align}
    \label{eq:T}
    \frac{dT}{dr} &= -\frac{m_0 g}{k}~\frac{1}{\partial_{V}f}~\frac{\Gamma_2-1}{\Gamma_2}\\
    \label{eq:rho}
    \frac{d\rho}{dr} &= -\frac{\rho m_0 g}{kT}~\frac{1}{\partial_{V}f}~\frac{1}{\Gamma_1} \\
    \label{eq:g}
    \frac{dg}{dr} &= 4\pi G\rho - \frac{2g}{r}
\end{align}

All quantities on the RHS are functions of $\rho$, $T$ and $g$ only (for a given value of $Y$). By imposing the central values $T_c$ and $\rho_c$ ($g_c$ must be null in any case), we now have a complete system the solution of which is $\displaystyle \begin{Bmatrix} T\,(r;Y,T_c,\rho_c) \\ \rho\,(r;Y,T_c,\rho_c) \\ g\,(r;Y,T_c,\rho_c) \end{Bmatrix}$. The resulting model is fully convective and contains an ionisation region whose properties are controlled by the triplet $(Y,T_c,\rho_c)$. Although this triplet seems the most intuitive, it is actually possible to parameterise the model by considering other triplets derived from these 3 quantities. Hereafter, we will consider instead the following dimensionless parameterisation:
\begin{align}
    Y_s ~&\equiv~ Y,\\
    \pa ~&\equiv~ \ln \left[\frac{\rho_c h^3}{2m_u(2\pi m_e kT_c)^{3/2}}\right], \\
    \pb ~&\equiv~ \frac{\chi_\textrm{H}}{kT_c}
\end{align} where we introduced the Planck constant $h$, the electron mass $m_e$ and the ionisation potential of hydrogen $\chi_\textrm{H}$. This particular choice and the associated notations will become clearer later on. At this point, one can see that this choice entirely constrains the $\Gamma_1$ value at the centre, the latter being only a function of $Y$, $2m_u/\rho\lambda_e^3$ (we keep in mind that $\lambda_e^{-2} = 2\pi m_e kT/h^2$), and $\chi_\textrm{H}/kT$ (cf. Eqs. \eqref{eq:Saha5}-\eqref{eq:yir} \& \eqref{eq:dTTf}-\eqref{eq:g1}). However, their impact on our model is much more significant than that.

\begin{figure*}
\centering
\includegraphics[width=\textwidth]{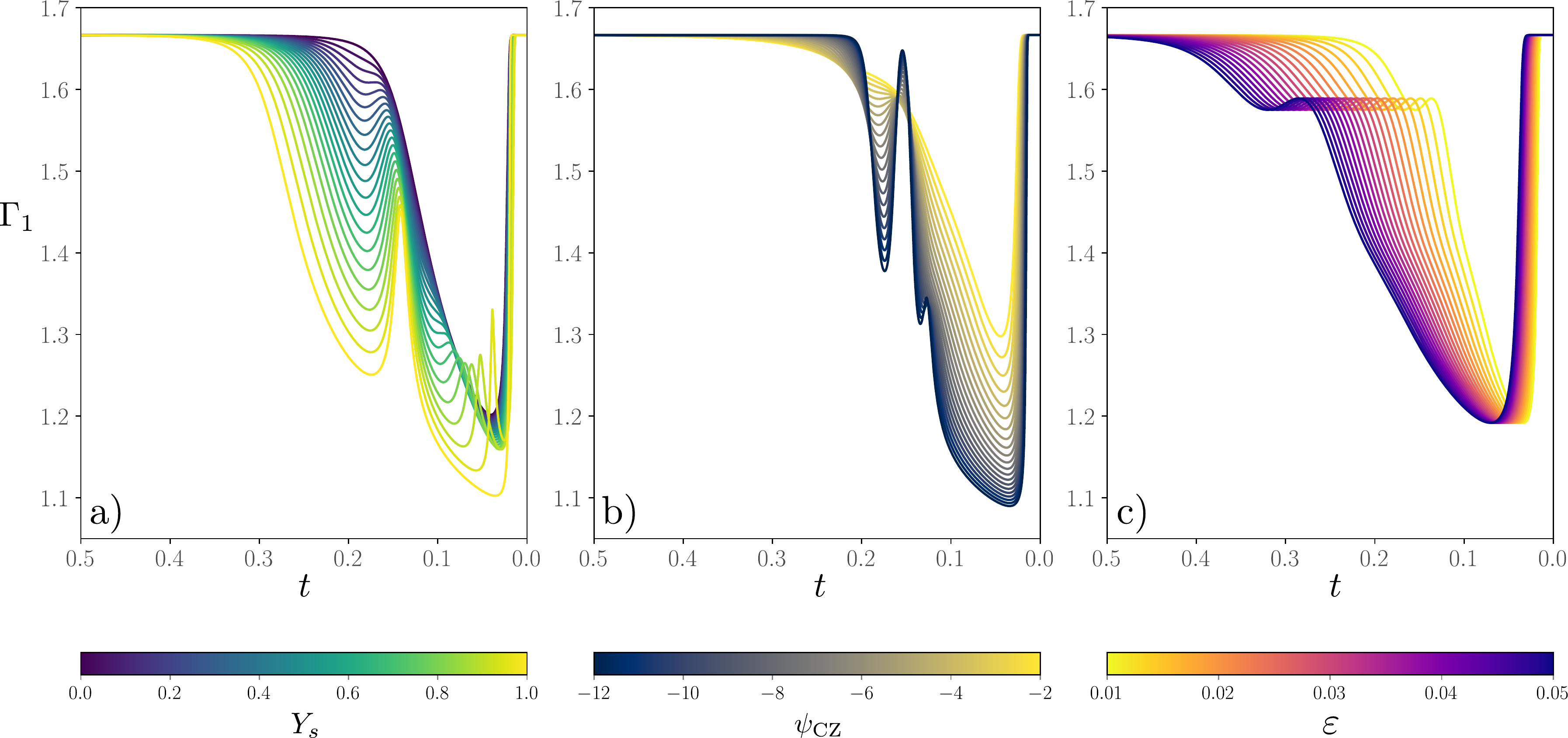}
\caption{$\Gamma_1(t)$ profiles obtained by considering various sets $(Y_s,\pa,\pb)$ around a reference $(Y_s^\odot,\pa^\odot,\pb^\odot) = (0.255,-5,0.015)$ that corresponds to a $\Gamma_1$ profile approaching that of the Sun (see Sect. \ref{CESTAM}). (a) Dependence of the profile with respect to $0\leq Y_s\leq 1$. (b) Dependence of the profile with respect to $-12\leq \pa \leq -2$. (c) Dependence of the profile with respect to $0.01\leq \pb \leq 0.05$. }
    \label{LinesG1.pdf}%
\end{figure*}

\paragraph{\emph{Physical interpretation of the parameters}}
\begin{itemize}
    \item Since $Y(r)$ is constant (see assumption 3), the helium amount $Y_s$ is clearly imposed at any location of the model, up to the surface symbolised by the index ``$s$''.
    \item The second parameter gets its notation from its close link to the electron degeneracy parameter $\psi = \mu_e/kT = \ln (n_e\lambda_e^3/2)$ in the ideal gas limit. In fact, one can show that:
    \begin{equation}
        \label{eq:psicenter}
        \psi \underset{\textrm{at centre}}{=} \pa - \frac{Y}{2} + \mathcal{O}(Y^2) 
    \end{equation}
    However, as long as $\Gamma_1 = 5/3$ (i.e. outside the ionisation region), $\rho \propto T^{3/2}$ which means that $\psi$ is preserved. Therefore, $\pa$ actually approximates the electron degeneracy parameter from the centre up to the ionisation region.
    \item The last parameter $\pb$ results from the ratio of two energy scales in the model. The first is the ionisation energy for the hydrogen ($13.6$ eV) and the second is the order of magnitude of the thermal energy at the centre of the model ($\sim$ keV), making this ratio generally rather small (hence the notation $\pb$). Through conservations, we just saw that $Y_s$ and $\pa$ actually reflect properties of the ionisation region; respectively its helium abundance and electron degeneracy. It is complicated to do so with $\varepsilon$ since $\chi_\textrm{H}/kT$ varies by several orders of magnitude in the model. We will however try to give an intuitive understanding of its impact on the model structure. The first thing to notice is that, if $\Gamma_1$ did not depend on $\chi_\textrm{H}/kT$, the latter would be equal to $5/3$ in the whole model. Indeed, since $\Gamma_1$ would depend only on $Y$ (which is constant) and $2m_u/\rho\lambda_e^3$ the value of which is fixed as long as $\rho\propto T^{3/2}$, the first adiabatic exponent would keep its central value, i.e. $5/3$. One can then see that the value of $\chi_\textrm{H}/kT$ at a given position alone determines whether $\Gamma_1$ deviates from the value $5/3$, i.e. if one enters the ionisation region starting from the model centre. In fact, it can be shown that the ionisation of the element $i$ in state $r$ takes place when: $\chi_i^r/kT \sim -\psi$. The role of $\pb$ can now be clearly identified: if one sets $\pb$ to be large (say of the order of $-\psi$), the ionisation will be complete close to the centre. In contrast, if $\varepsilon$ is small before $-\psi$, it will take until $T\ll T_c$ for the ionisation to end and the region will be shifted closer to the surface. Thus, despite not being explicitly linked to any quantity of the ionisation zone, $\varepsilon$ controls the position of the region.
\end{itemize}

Having given a physical interpretation for the parameters, we will now try to provide a geometrical interpretation of the impact of each parameter regarding the $\Gamma_1$ profile. We will rely on Fig. \ref{LinesG1.pdf}, which shows the $\Gamma_1(t)$ (with $t = \tau/\tau_0$, the normalised acoustic depth) profiles resulting from multiples parameter sets $(Y_s,\pa,\pb)$ by varying them one by one around a reference value.

\paragraph{\emph{Geometrical interpretation of the parameters regarding the $\Gamma_1$ profile}}
\begin{itemize}
    \item We gave a representation of $\Gamma_1(t)$ for all possible values of $Y_s$ in panel (a) of Fig. \ref{LinesG1.pdf}, including such extreme cases as a full pure hydrogen or helium stars. As expected, we go from a profile showing only the hydrogen well ($Y_s = 0$) to one showing only the two helium wells ($Y_s = 1$), passing through profiles that account for the $3$ components ($0<Y_s<1$). Clearly, increasing $Y_s$ has the effect of extending both helium wells while reducing the hydrogen one. 
    \item The effects of modifying $\pa$ can be seen in panel (b) of Fig. \ref{LinesG1.pdf}. The range under consideration corresponds to typical solar degeneracy values ranging from the surface ($-12 :$ very low degeneracy) to the centre ($-2 :$ high degeneracy). Once more, the impact is clearly visible: a high degeneracy tends to spread out the ionisation wells making them indistinguishable when $\pa \rightarrow -2$. In the opposite situation, low degeneracies in the ionisation region result in much deeper and localised wells. It is even possible to distinguish the contributions of the hydrogen and the first helium ionisations when $\pa$ gets closer to $-12$. In fact, this last point is consistent with what has been seen in Fig. \ref{MapG1.pdf} by noticing that high degeneracies correspond to relations of the form $\ln \rho(T) = C+(3/2)\ln T$ with high $C$ values. Thus, structures with high $\psi_\textrm{CZ}$ correspond to $\rho(T)$ that are located in the upper part of Fig. \ref{MapG1.pdf}, where the ionisation regions tend to overlap. The opposite reasoning holds just as well.
    \item The impact of $\pb$ is illustrated in panel (c) of Fig. \ref{LinesG1.pdf}. As expected from the previous paragraph, varying its value from $0.01$ to $0.05$ results in a shift of the ionisation region. However this change differs between each well, the one of HeII being more affected than the hydrogen one. In order of magnitude, this effect can be explained by the dependence of the normalised acoustic depth on temperature close to the surface: $t \propto \sqrt{T/T_c} \propto \sqrt{\pb T}$. Thus, for a given ionisation temperature $T_\textrm{ion}$, the impact of $\pb$ will become more noticeable on the associated depth $t_\textrm{ion}$ as $T_\textrm{ion}$ is high. This reasoning also explains why the shift does not seem linear with respect to $\pb$ for a given well (this is particularly visible for the HeII well); the latter actually varies as $\sqrt{\pb}$.
\end{itemize}

   \begin{figure*}[ht!]
   \centering
   \includegraphics[width=\textwidth]{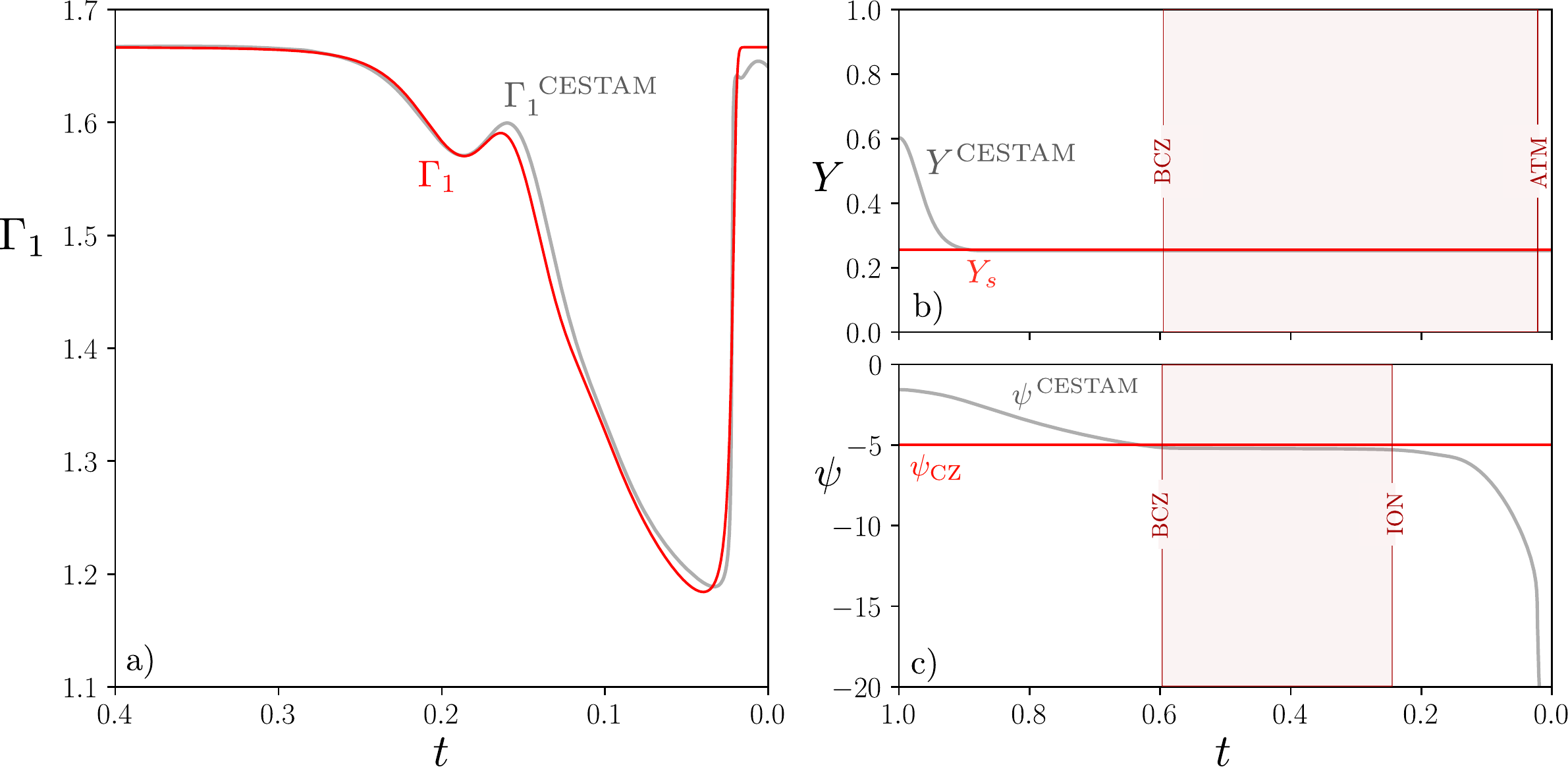}
      \caption{(a) Comparison of $\Gamma_1^{~\textrm{CESTAM}}(t)$ extracted from a CESTAM solar model (in grey) and $\Gamma_1(t;Y_s^\odot,\pa^\odot,\pb^\odot)$ obtained from our model (in red). (b) Comparison of $Y^{\textrm{CESTAM}}(t)$ extracted from the same model with $Y_s^\odot$. (c) Comparison of $\psi^{\textrm{CESTAM}}(t)$ with $\pa^\odot$. In the last two panels, regions where the curves are expected to correspond are shaded in red. }
         \label{YandPsi.pdf}
   \end{figure*}

Regarding Fig. \ref{LinesG1.pdf}, an interesting point of the model defined with $(Y_s,\pa,\pb)$ is the relative independence of the impact of each parameter. Indeed, by treating each well as a distribution, all parameters seem to impact a specific moment. For instance, $Y_s$ seems directly related to the distribution \textit{area}, and does not impact much the shapes of the wells nor the position (for reasonable values of $Y_s$). $\pa$ is clearly related to the \textit{dispersion} of the distribution and $\pb$ to its \textit{position}. Such distinct impacts on the structure are encouraging for subsequent parameter retrieval as one can expect to avoid degeneracy biases.

\subsection{Comparison with a realistic stellar model \label{CESTAM}}

The first and most important point to address is to verify to what extent the model described in this article can approximate the ionisation region structure of a realistic stellar model. To that end, we compare in panel (a) of Fig. \ref{YandPsi.pdf} $\Gamma_1(t)$ profiles extracted from a CESTAM solar model and from our model for a particular parameter set $(Y_s^\odot,\pa^\odot,\pb^\odot)$. These values have been chosen so that 
\begin{equation}
    \label{eq:intdG1}
    \int_0^1\left(\frac{\delta_t\Gamma_1}{\Gamma_1}\right)^2\,dt = \int_0^1\frac{\left(\Gamma_1^{~\textrm{CESTAM}}(t)-\Gamma_1(t;Y_s,\pa,\pb)\right)^2}{\Gamma_1^{~2}}\,dt
\end{equation} is minimal. One can see that the behaviour of the realistic model can be very well reproduced, the curves differing mainly in only 2 regions. A first difference is located in the first helium ionisation region ($0.12<t<0.18$), where the depth of the well is slightly overestimated. This mainly results from the assumption made in Sect. \ref{EQ}, where we assumed that $\Bar{z}\sim 1$ in the helium ionisation region and thus underestimated the number of electrons in the HeI region. This causes the ionisation to start a little deeper in the model and therefore to be more localised. By reducing the dispersion of the HeI well, the assumption made its depth slightly greater. As expected, another difference can be seen in the atmospheric region ($t<0.03$). Because the ionisation has not yet started, our $\Gamma_1$ still has the value $5/3$, which is obviously not the case in a realistic model containing an atmospheric modelling. The overall profile shape is nevertheless well reproduced in the ionisation region, especially considering that only $3$ parameters need to be adjusted.\newline

A second point to be clarified is the exact meaning of the adjusted values $(Y_s^\odot,\pa^\odot,\pb^\odot)$ regarding the realistic model. Having the minimal differences $(\delta_t\Gamma_1/\Gamma_1)^2$, the model parameterised by $(Y_s^\odot,\pa^\odot,\pb^\odot)$ is interpreted as being the closest to the CESTAM model from the perspective of a particular structural aspect \footnote{This aspect will be discussed in much more detail in a forthcoming paper but one can already relate it to the ``visible'' part of the structure via the frequency shift considering Eq.~\eqref{eq:kernel_HG07}}. In this sense, if our modelling and its interpretation are correct, we can expect our parameters to approach the quantities they refer to in the realistic model. However, in previous paragraphs we gave an understanding of these parameters in \textit{our} model but a realistic model only locally verifies the assumptions used to derive our structure. In particular, the relation between $P$ and $\rho$ will be described by $\displaystyle \gamma\equiv \frac{d\ln P}{d\ln\rho}$ rather than $\Gamma_1$, the two differing outside the isentropic region. Accordingly, we expect the adjusted values to approximate the CESTAM model quantities where assumptions 1 to 4 are verified, i.e. in the CZ. For the helium abundance $Y_s$, since $Y_s = Y(t)$ at any position of our model, the value $Y_s^\odot$ should approximate $Y^\textrm{CESTAM}(t)$ in the whole CZ up to the atmosphere. In the case of $\pa$, we saw that the approximation remains valid as long as $\Gamma_1 = 5/3$, so we expect $\pa^\odot$ to approach $\psi^\textrm{CESTAM}(t)$ in the CZ part located below the ionisation region. As mentioned, since $\pb$ does not relate to any particular quantity of the CZ, it is difficult to deduce more than the position of the ionisation region. These approximations can be summarised as follows:
\begin{align}
    \label{eq:Yapp}
    Y_s^\odot &\simeq Y^\textrm{CESTAM}(t) \qquad\textrm{for \quad BCZ $<t<$ ATM} \\ 
    \label{eq:psiapp}
    \pa^\odot &\simeq \psi^\textrm{CESTAM}(t) \qquad \textrm{for \quad BCZ $<t<$ ION}
\end{align} with BCZ, ATM and ION designating respectively the base of the convective zone, the base of the atmosphere and the end of the ionisation region. We illustrate this point in panels (b) and (c) of Fig. \ref{YandPsi.pdf} by representing both $Y^\textrm{CESTAM}(t)$ and $\psi^\textrm{CESTAM}(t)$, as well as the estimates $Y_s^\odot$ and $\pa^\odot$ obtained through the minimisation of Eq.~\eqref{eq:intdG1}. We also highlight the regions were the quantities are expected to correspond to each other based on Eqs.~\eqref{eq:Yapp} and~\eqref{eq:psiapp}. The values thus obtained $(Y_s^\odot,\pa^\odot,\pb^\odot) = (0.255,-5,0.015)$ are consistent with the CESTAM quantities in the associated regions (see panels (b) \& (c) of Fig. \ref{YandPsi.pdf} and Table \ref{tab:summary}).\newline

\begin{table}[!ht]
    \begin{center}
    \caption{Inferred electron degeneracy parameter and helium abundance and actual CESTAM quantities}
    \begin{tabular}{ccc}
        \hline
        \hline~\\[-3mm]
        \textbf{Quantity} & \textbf{Inferred value} & \textbf{CESTAM} \\
        \hline~\\[-2mm]
        $Y$ & $0.255$ & $0.253$ \\[1mm]
        $\psi$ & $-5.00-0.13=-5.13^{~(*)}$ & $-5.19$ \\[1mm]
        \hline~\\[-1mm]
    \end{tabular}
    \label{tab:summary}
    \end{center}
    \footnotesize
    $^{~(*)}$ : Note that our $\psi$ estimate results from the combination $\pa-Y_s/2$ in accordance with Eq.~\eqref{eq:psicenter}.
    \normalsize
\end{table}

It is risky to be quantitative about this result since there are no obvious values with which the difference can be compared; in particular the relative difference does not seem relevant. Nevertheless, this example strongly suggests that the modelling defined here allows us to define structures that are close to realistic models (in the sense of criteria \eqref{eq:intdG1}) and to also obtain similar helium abundances and electron degeneracies in the regions of interest.

\section{Analysis of first adiabatic exponent perturbations\label{DISCUSSION}} 

\begin{figure*}
\centering
\includegraphics[width=\textwidth]{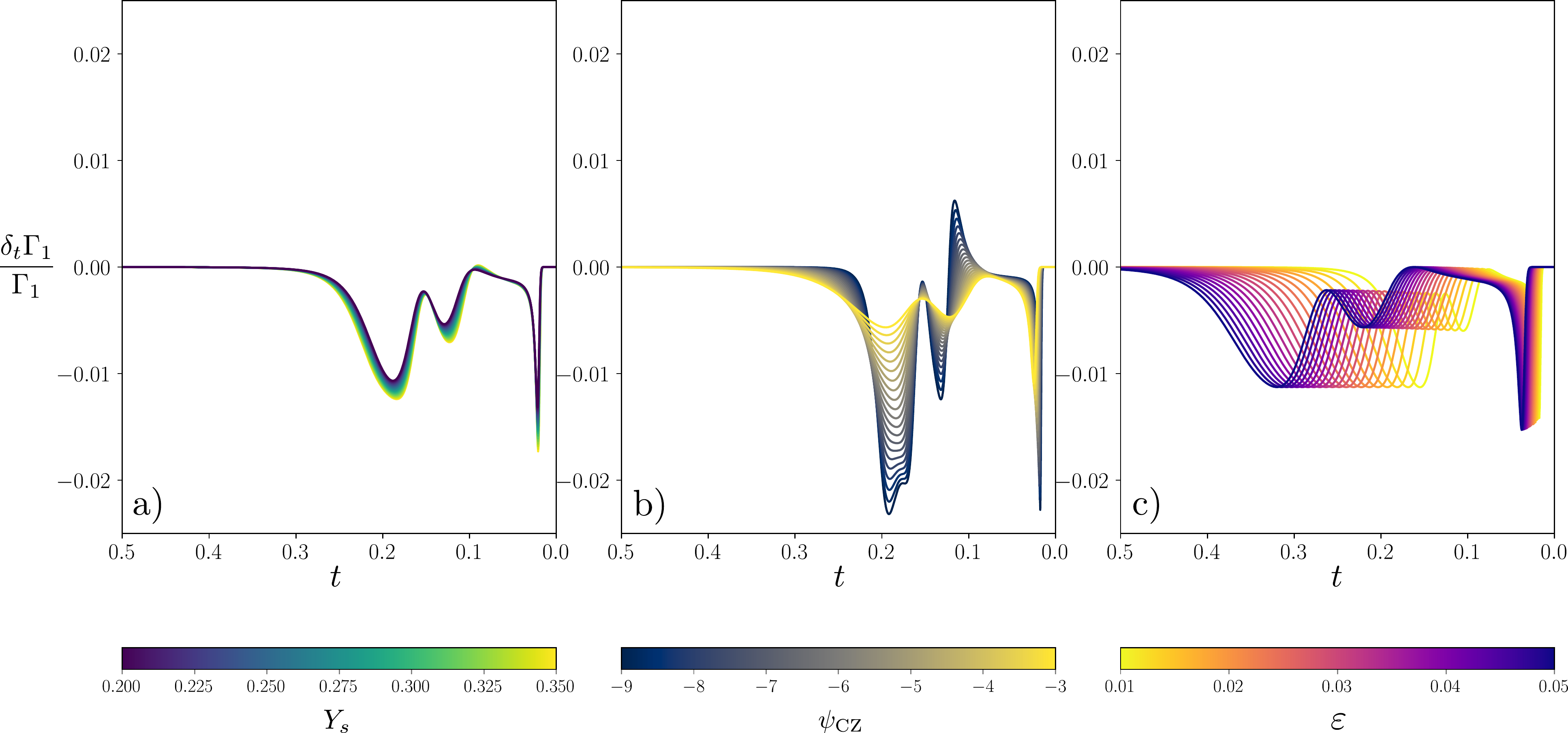}
\caption{Profile differences $\delta_t\Gamma_1/\Gamma_1 = \left[\Gamma_1(t;Y_s+\delta Y_s,\pa,\pb)-\Gamma_1(t;Y_s,\pa,\pb)\right]/\Gamma_1$ obtained for a fixed value of $\delta Y_s = 0.05$ and  various sets $(Y_s,\pa,\pb)$ (note that the values used here are not the same as those in Fig. \ref{LinesG1.pdf}) around the reference $(Y_s^\odot,\pa^\odot,\pb^\odot) = (0.255,-5,0.015)$ as function of the normalised acoustic depth $t$. (a) Dependence with respect to $0.2\leq Y_s\leq 0.35$. (b) Dependence with respect to $-9\leq \pa \leq -3$. (c) Dependence with respect to $0.01\leq \pb \leq 0.05$.}
    \label{dYLinesG1.pdf}%
\end{figure*}

So far, we focused on introducing a physical model of the ionisation region that depends on only a few parameters in order to study its properties. However, as can be seen from Eq. \eqref{eq:kernel_HG07}, the analysis of frequency shifts relies more on the modelling of structural perturbations than on the structure itself. We will therefore investigate what our model predicts as a perturbation caused by a change in surface helium abundance compared with the ad-hoc profiles used in previous papers (see Fig. \ref{AllGlitches.pdf}). The advantage of having modelled the structure is that we can easily reduce it to the study of a perturbation of helium abundance by considering the following difference between profiles:
\begin{equation}
    \label{eq:dG1_dY}
    \frac{\delta_t\Gamma_1}{\Gamma_1} = \frac{\Gamma_1(t;Y_s+\delta Y_s,\pa,\pb)-\Gamma_1(t;Y_s,\pa,\pb)}{\Gamma_1}
\end{equation}

The analysis of this perturbation is carried out in Fig. \ref{dYLinesG1.pdf}. An important aspect to bear in mind is the number of variables on which the above function depends. The perturbation will naturally depend on the helium difference $\delta Y_s$ but also on the point, $(Y_s,\pa,\pb)$, around which the differences are calculated, thus leading to 4 dimensions to explore. Our analysis of the three-dimensional space required for the structure was obviously already incomplete in Fig. \ref{LinesG1.pdf}, and it follows that the perturbation analysis will necessarily be even more superficial. Figure \ref{dYLinesG1.pdf} provides plots for various reference points, $(Y_s,\pa,\pb)$, but for a fixed difference $\delta Y_s = 0.05$. Nevertheless, it is straightforward to see by linearising Eq.~\eqref{eq:dG1_dY} that a difference caused by $\delta Y_s' = \alpha\times\delta Y_s$ would be relatively similar to the profile $\alpha\times \delta_t\Gamma_1/\Gamma_1$. In this sense, Fig.~\ref{dYLinesG1.pdf} still provides a good representation of what a change in helium abundance might possibly cause. We see in panel (a) that all the perturbed profiles roughly overlap, which reflects a relative independence from the helium amount chosen as a reference (the range considered, $0.2<Y_s<0.35$, being representative of most realistic stellar helium abundances). The shape is quite similar to what is expected from a helium variation: each helium ionisation region contributes to a Gaussian-like well ($t\sim0.13$ \& $t\sim0.19$), the second one being more pronounced. The hydrogen contribution is less intuitive, resulting in a very localised peak ($t\sim0.03$, present on every panel) at the beginning of ionisation. The latter reflects more of a shift of the hydrogen ionisation region (visible in panel (a)) of Fig. \ref{LinesG1.pdf}) caused by the helium change than a real drop in the well. Nevertheless, it should be noted that this component is clearly noticeable in terms of amplitude.

\begin{figure*}
\centering
\includegraphics[width=\textwidth]{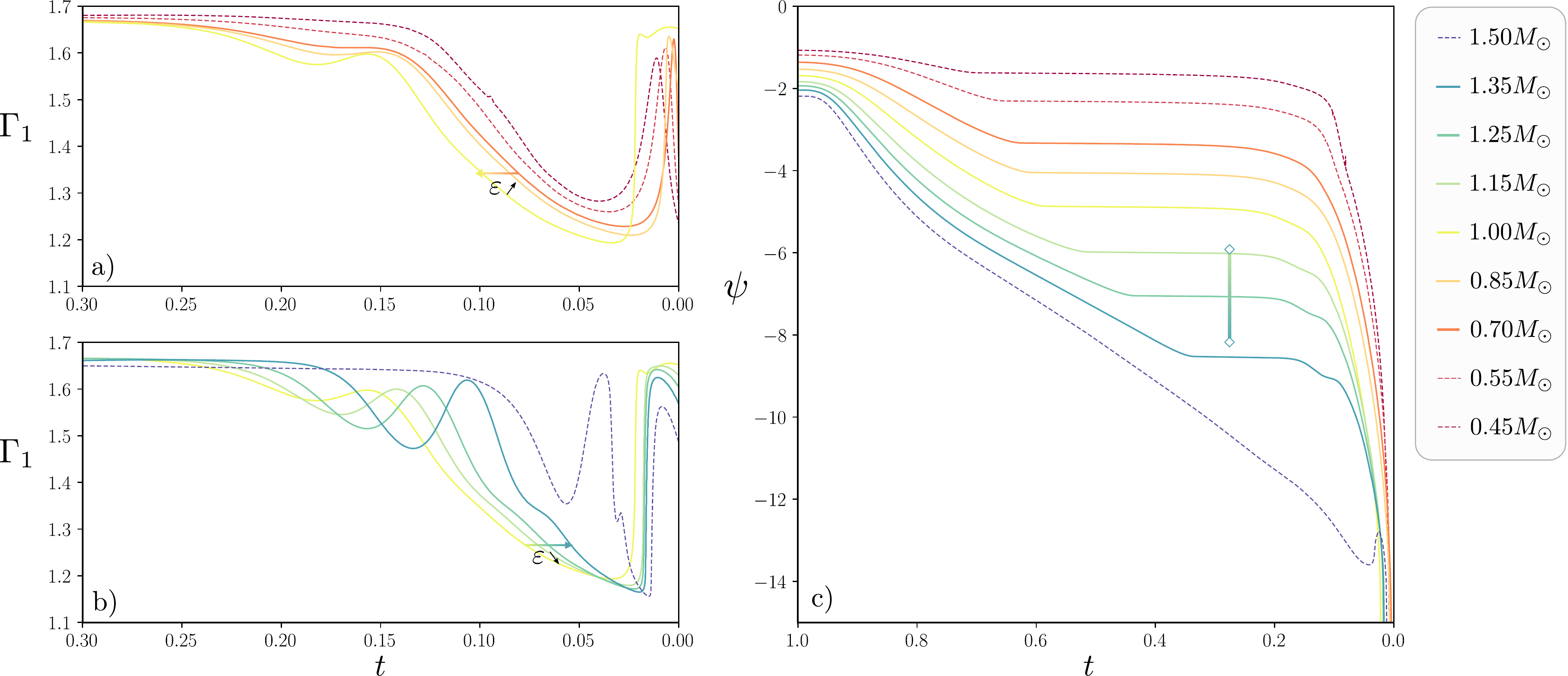}
\caption{$\Gamma_1(t)$ (both panels (a) \& (b) for clarity) and $\psi(t)$ (panel (c)) profiles extracted from CESTAM early main sequence models with masses ranging from $0.7 M_\odot$ to $1.35M_\odot$ (and from $0.45 M_\odot$ to $1.5M_\odot$ as an indication in dashes). All models share the same composition which is the solar one as well as the same evolutionary state: $X_c = 0.6$. The qualitative variations of $\varepsilon$ with mass are represented as well as the $\psi$ span corresponding to the mass uncertainty of HD 52265 taken from \cite{Lebreton2014}.}
    \label{AllM.pdf}%
\end{figure*}

Things get less intuitive in panel (b). In order to provide realistic values for this rather uncommon parameter, we considered $-9<\pa<-3$ as the electron degeneracy range which corresponds to typical values inside solar-like oscillators (a justification of this point is provided in Fig. \ref{AllM.pdf}). This time the shape of the perturbation caused by a fixed change in the helium amount is very sensitive to the electron degeneracy value taken as a reference, and highlights how counter-intuitive differences of profiles can be. At higher electron degeneracy levels the perturbation looks like a more spread out version of the one visible in panel (a), however its behaviour becomes more complex as the degeneracy decreases. Both helium wells lose their Gaussian aspect: the first one gets closer to a ``heartbeat'' shape thus becoming clearly positive (probably under the influence of the hydrogen well) while the second becomes progressively bi-modal. In contrast, the results in panel (c) could have been guessed from Fig. \ref{LinesG1.pdf}. The $\pb$ value shifts both reference and perturbed profiles, thus resulting in a scaled version of the perturbation shown in panel (a).\newline

\begin{figure}
\centering
\includegraphics[width=8.5cm]{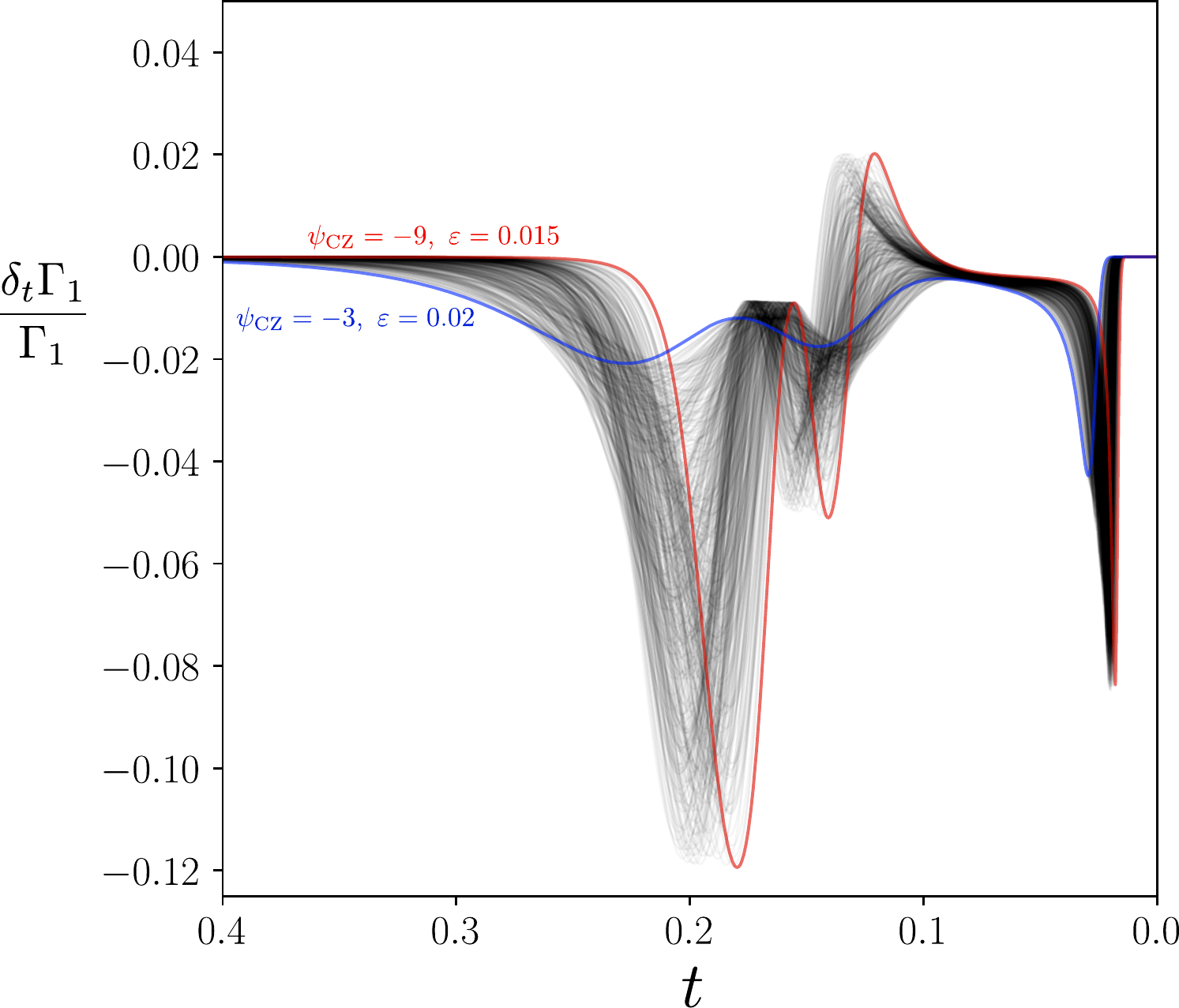}
\caption{Profile differences $\delta_t\Gamma_1/\Gamma_1$ obtained once again for a fixed value of $\delta Y_s = 0.25$ but from a pure hydrogen reference model $Y_s = 0$. The $500$ profile differences in this plot were obtained by randomly drawing the remaining two reference quantities $(\pa, \pb)$ in the two dimensional box $[-9, -3]\times[0.015,0.02]$ (note that the $\varepsilon$ interval has been reduced compared to Fig.~\ref{dYLinesG1.pdf} for clarity reasons). Two opposite corners of this box have been highlighted in red and blue.}
    \label{random500_dYLinesG1.pdf}%
\end{figure}

As a consequence, Fig. \ref{dYLinesG1.pdf} first warns us about the issues inherent to calibration methods already mentioned in Section \ref{REVIEW}. Even if it can be established that the choice of the helium amount $Y_s$ taken as reference does not ultimately matter much, Fig. \ref{dYLinesG1.pdf} illustrates the diversity of $\delta_t\Gamma_1/\Gamma_1$ profiles corresponding to the same helium difference by simply considering different values of $\pa$ and $\pb$. Reusing the notations introduced in Section \ref{REVIEW}, it is clear that any change in a component of $\Vec{\theta}_\star$ that may impact the best-fitting couple $(\pa, \pb) \subset \Vec{\theta}_p$ will lead to substantially different estimates of $\delta Y$. In this respect, a parameter with a particularly strong impact will be given in the forthcoming paragraphs, thus further distancing us from a unique relation of the form $\Vec{\theta}_p(\delta Y)$ as assumed in Eq.~\eqref{eq:calibration}. Furthermore, it should be added that the particular case of a perturbation around the reference value $Y_s=0$ (as considered in HG07) is no exception.  Although we only varied the two remaining parameters $\pa$ and $\pb$, Fig.~\ref{random500_dYLinesG1.pdf} shows the many possibilities for $\delta_t\Gamma_1/\Gamma_1$ even when considering an identical helium difference $\delta Y_s = 0.25$ from a pure hydrogen model $Y_s = 0$. The spread of these randomly drown curves provides an idea of the potential dispersion of the perturbed profile at any point of the structure even under these ``favourable'' conditions. 

Secondly, Figs. \ref{dYLinesG1.pdf} \& \ref{random500_dYLinesG1.pdf} underline how accustomed we are in conceiving the structural perturbations caused by a variation in helium under solar conditions of electron degeneracy. We may note here the added value of having introduced a model physically; it would have been complicated to imagine and then parameterise such types of perturbations by simply observing a $\Gamma_1$ profile in a realistic model. Accordingly, we may question the validity of using ad hoc profiles to approximate the perturbation caused by a change in helium abundance, and this particularly at low electron degeneracy (whose domain of relevance will be discussed in the following). Indeed, it seems likely that adjusting functions whose form is inappropriate for these complex profiles may result in inconsistencies regarding their parameterisations. \newline

\begin{figure*}
\centering
\includegraphics[width=\textwidth]{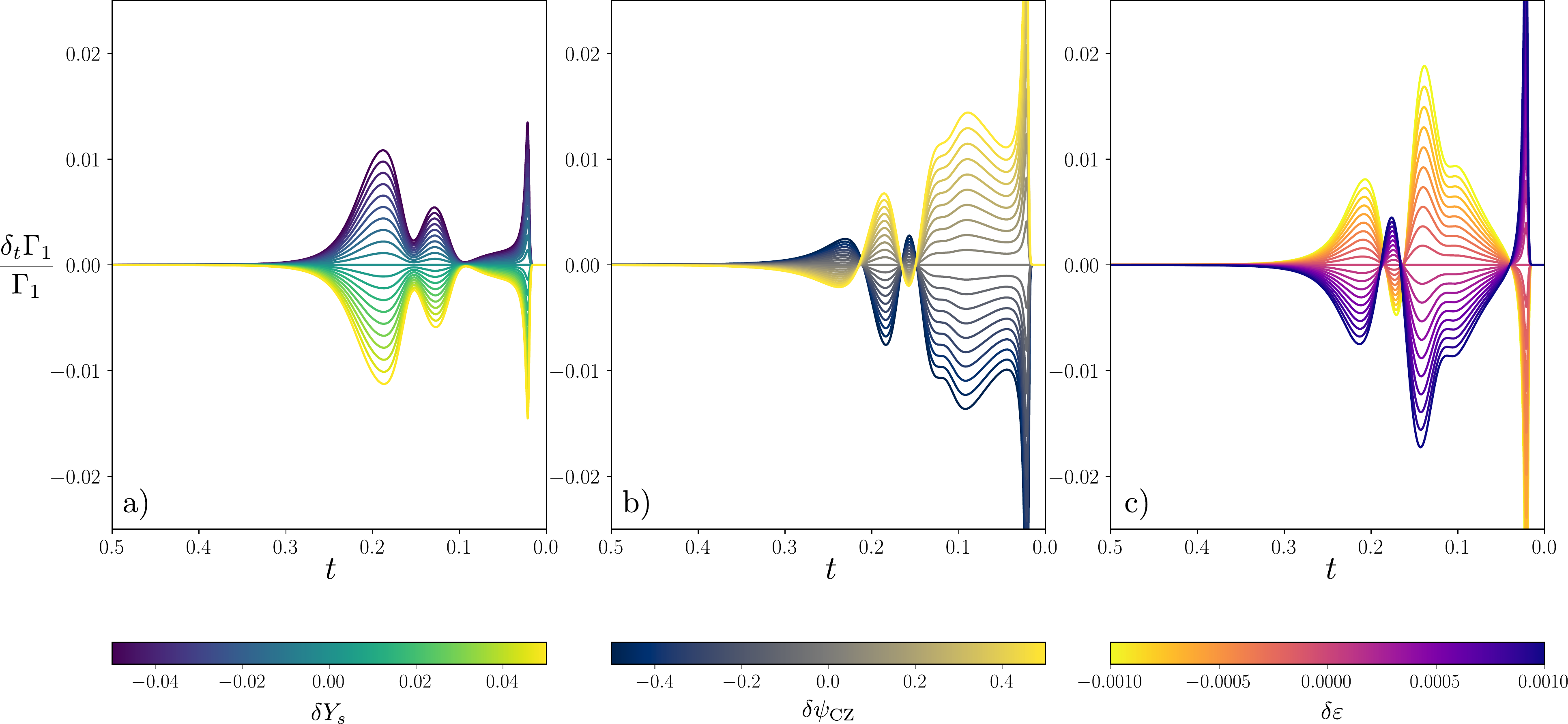}
\caption{Profile differences $\delta_t\Gamma_1/\Gamma_1 = \left[\Gamma_1(t;Y_s+\delta Y_s,\pa+\delta\pa,\pb+\delta\pb)-\Gamma_1(t;Y_s,\pa,\pb)\right]/\Gamma_1$ obtained for a fixed reference parameter set $(Y_s,\pa,\pb) = (Y_s^\odot,\pa^\odot,\pb^\odot)$ and by considering various sets $(\delta Y_s,\delta\pa,\delta\pb)$. (a) Dependence with respect to $(\delta Y_s,0,0)$ for $-0.05\leq \delta Y_s\leq 0.05$. (b) Dependence with respect to $(0,\delta\pa,0)$ for $-0.5\leq \delta\pa \leq 0.5$. (c) Dependence with respect to $(0,0,\delta\pb)$ for $-0.001\leq \delta\pb \leq 0.001$. }
    \label{dLinesG1.pdf}%
\end{figure*}

Up to this point, we have only dealt with a given helium difference, in particular to see how the model presented in this paper agrees or contrasts with the types of perturbations for ionisation regions from previous studies. However, it is clear that Eq. \eqref{eq:dG1_dY} extends to the much more general difference: 
\begin{equation}
    \label{eq:dG1_dAll}
    \frac{\delta_t\Gamma_1}{\Gamma_1} = \frac{\Gamma_1(t;Y_s+\delta Y_s,\pa+\delta\pa,\pb+\delta\pb)-\Gamma_1(t;Y_s,\pa,\pb)}{\Gamma_1}
\end{equation}
Indeed, when applied to frequency shift analysis, it seems unlikely that the model differs from its target only in its helium abundance. Therefore, in the following, we propose to study the structural impact of a $\delta\pa$ or a $\delta\pb$ difference. It is obvious that the problems mentioned above in terms of the number of dependencies will only be further amplified. For this reason, we fixed the reference set of parameters by taking for example the values $(Y_s^\odot,\pa^\odot,\pb^\odot)$ and varied one by one the differences $\delta Y_s$, $\delta\pa$ and $\delta\pb$. The resulting profiles are shown in Fig. \ref{dLinesG1.pdf}.

As expected, the variation observed in panel (a) is very similar to the one in panel (a) of Fig. \ref{dYLinesG1.pdf} scaled by a factor $\alpha = \delta Y_s/0.05$. The other two perturbations, however, take different shapes and notably make the wells unrecognisable. A change in the electron degeneracy (panel (b)) has a strong impact on the hydrogen ionisation structure, especially at a point ($t\sim0.09$) where a change in helium has almost no effect. The ionisation region shift is also enhanced while the second helium ionisation contributes to a weaker and more localised well. The variation in the position $\delta\pb$ of the ionisation region is also fairly counter-intuitive. Indeed, panel (c) illustrates a contribution that reaches its maximum between the two helium wells ($t\sim0.15$), the one corresponding to the peak in the $\Gamma_1(t)$ profile.

A first observation to highlight from Fig. \ref{dLinesG1.pdf} is how elaborate the ionisation perturbation profile becomes in the general case. It should be kept in mind that a structural perturbation from a reference $(Y_s^\odot,\pa^\odot,\pb^\odot)$ can be approximated as a \textit{combination} of profiles composing Fig. \ref{dLinesG1.pdf}. Indeed, from Eq.~\eqref{eq:dG1_dAll}:
\begin{equation}
    \label{eq:dG1lin}
    \frac{\delta_t\Gamma_1}{\Gamma_1} \simeq \delta Y_s \left.\frac{\partial \ln \Gamma_1}{\partial Y_s}\right|_\odot + \delta \pa \left.\frac{\partial \ln \Gamma_1}{\partial \pa}\right|_\odot + \delta \pb \left.\frac{\partial \ln \Gamma_1}{\partial \pb}\right|_\odot
\end{equation} with e.g.,
\begin{equation}
    \left.\frac{\partial \ln \Gamma_1}{\partial Y_s}\right|_\odot \equiv \lim_{\delta Y_s \rightarrow 0} \frac{\Gamma_1(t;Y_s^\odot+\delta Y_s,\pa^\odot,\pb^\odot)-\Gamma_1(t;Y_s^\odot,\pa^\odot,\pb^\odot)}{\Gamma_1 \delta Y_s}
\end{equation} being the structural perturbation studied in panel (a) within a multiplicative factor. With regard to Fig. \ref{dLinesG1.pdf}, the result of the linear combination \eqref{eq:dG1lin} can become highly complex and is unlikely to have anything in common with the functions generally used to fit $\delta_t\Gamma_1/\Gamma_1$. Combined with the issues already raised about calibration, this tends to suggest that the study of the ionisation glitch based on ad hoc perturbation profiles in order to calibrate them afterwards might lead to fairly inconsistent results.

Also, we have tried to represent in the figure structural disturbances that are similar in order of magnitude. This allows us to qualitatively relate the values of the different parameters. In this sense, we see that perturbation with $\delta Y_s = 0.1$, $\delta\pa = 1$ or $\delta\pb = 0.002$  have a comparable structural impact in amplitude and hence on the frequencies. However, although the first one nearly spans the entire range of relevant values (the difference in helium between two stars can hardly exceed $0.1$), this is not the case for the two others. In fact, regarding the $\pb$ parameter, this change is actually very small: two consecutive curves in panel (c) of Fig. \ref{LinesG1.pdf} are separated by exactly $\delta\pb = 0.002$ which is the entire span for the perturbation represented in Fig. \ref{dLinesG1.pdf}! This last point suggests a very high dependence of the perturbation on a shift of the ionisation region. As mentioned above, it is interesting to note that the glitch signature would come from the $\Gamma_1$ peak rather than from a well if this contribution were to dominate. This is consistent with observations already made in previous studies \citep{Broomhall2014,Verma2014b}, although it provides an alternative explanation.

To put into perspective the significance of a difference in the $\pa$ parameter, it can be interesting to relate it to more intuitive quantities like fundamental parameters. Since the electron degeneracy $\psi$ is known to fluctuate with the mass of the star \citep[e.g.][]{Hayashi1962}, we represented in Fig. \ref{AllM.pdf} various $\psi(t)$ profiles (along with the corresponding $\Gamma_1(t)$ profiles) obtained from CESTAM models of different masses. The latter cover a wide range of values from $0.7M_\odot$ to $1.35M_\odot$ appropriate for solar-like pulsators. All the models also share the same composition (which is the solar one) and subsequently the same helium abundance as well as the same evolutionary state ($X_c = 0.6$). Clearly, despite having an identical composition, these models show fairly distinct $\Gamma_1(t)$ profiles in the ionisation region. In fact, these curves show a fairly similar behaviour to those shown in panel (b) of Fig. \ref{LinesG1.pdf} (which can be extrapolated to the wider range $0.45M_\odot<M<1.5M_\odot$ presented in dashes as an indication). Thus, at small masses, the wells seem to merge and then split at higher masses making in particular the HeI well visible. In addition to this effect, an expansion and contraction of the profiles is visible depending on the mass range considered ($<1 M_\odot$ or $>1M_\odot$, cf. arrows shown in panels (a) \& (b)). Looking at the similarities with Fig. \ref{LinesG1.pdf}, we interpret these changes as manifestations of variations in $\varepsilon$ and the electron degeneracy just below the ionisation region. Although the variations of $\varepsilon$ can only be evaluated qualitatively (remember that this quantity is only defined in entirely convective models), this is not the case for the electron degeneracy. Indeed, $\psi(t)$ profiles typically undergo little to no variation in convection zones as shown by the plateaus in the curves in panel (c),  thus making it easy to identify $\pa$. The range of $\pa$ values covered by these plateaus corresponds to the range studied in Fig.~\ref{dYLinesG1.pdf} ($\sim-3$ for $0.7M_\odot$ stars and $\sim-9$ for the $1.35M_\odot$ model). It is then possible to seek to characterise the significance of the change $\delta\pa = 1$ mentioned above. To provide an order of magnitude, we represented in panel (c) the $\psi$ span corresponding to the mass uncertainty ($1.14<M_*<1.32$) of HD 52265 taken from \cite{Lebreton2014}. The latter provides an example of a seismically
determined mass uncertainty taking into account a large number
of biases as discussed in Section \ref{INTRO}. The corresponding extent in terms of electron degeneracy is about $\delta\pa \sim 2$, which is 2 times larger than the range considered in the panel (b) of Fig. \ref{dLinesG1.pdf} and thus larger than any possible variation in helium abundance. Note that this range in $\pa$ depends on the mass; such a mass difference would correspond to a lower $\delta\pa$ at $0.7M_\odot$ and would instead increase at higher masses. In this regard, a mass difference between two models\footnote{Here, one may note that a change in mass at a \textit{given evolutionary state} has been studied. The impact of an age difference at a given mass has also been examined, which leads to smaller changes in both $\pa$ and $\pb$ under the constraints provided by \cite{Lebreton2014}. The only substantial variation in the $\Gamma_1$ profile occurs in the $1.5 M_\odot$ model, with the appearance of a convective surface zone for $X_c \leq 0.4$.} could contribute to a perturbation of the structure of the ionisation region as large as (if not larger than) that of helium.

Finally, to further echo the point made in Fig.~\ref{dYLinesG1.pdf} about calibration, the mass taken as a reference, $M_\star$, provides a representative example of a component of $\Vec{\theta_\star}$ that impacts the $(\pa,\pb)$ pair as already observed by \cite{Verma2014b}. In this article, the authors proposed to use it by considering different calibrations to determine the helium abundance of stars of distinct masses. However, besides the fact that this presupposes a relatively good knowledge of the star in question before being able to estimate its helium abundance, we have seen that stars whose masses are already seismically constrained can possess significantly different degeneracy levels in the convective region. Using the example of HD 52265, the question then arises whether it is possible to obtain consistent results using calibration and being able to consider both models with $\pa \sim 6$ or $\pa \sim 8$, which implies perturbation profiles that differ in both form and magnitude for a fixed difference in helium amount (see panel (b)) of Fig.~\ref{dYLinesG1.pdf}). This statement does not apply exclusively to the mass, $M_\star$, but can reasonably be extended to any fundamental parameters or physical processes (i.e. the other components of $\Vec{\theta}_\star$) that may affect the average density in the convection zone (related to $\pa$) or the relative acoustic depth of the ionisation region (linked to $\pb$).\newline

Besides this last remark, by comparing the structural perturbations in Fig. \ref{dLinesG1.pdf}, it seems likely that the component caused by a helium change between two stars is not the only one to contribute. In fact, it could even be dominated by a change in electron degeneracy or in the ionisation region position. From the perspective of frequency shifts, it thus seems incorrect to assimilate the ionisation glitch as a repercussion of a helium difference only in order to infer its value. In this light, it appears essential to look at forms of frequency shifts that can take into account these additional dependencies, thus relying on physical models as proposed in this paper.

\section{Conclusion\label{CONCLUSION}}

Determining abundances can be subject to many biases in its classical approach, which we intend to overcome by exploiting the ionisation glitch. However, although progressively becoming more sophisticated, the glitch-based approach faces problems inherent to its current modelling such as the need for calibration by realistic stellar models.

To address these problems, a physical model of the ionisation region is proposed here, explicitly involving the parameters of interest, such as the surface helium abundance $Y_s$, and which can be generalised to more elaborate compositions. In the case of a hydrogen-helium mixture, the model involves 3 parameters and highlights the importance of a characterisation of the electron degeneracy state in the convective zone $\pa$ as well as the position of the ionisation region here controlled by $\pb$. While it is well known that the abundance of surface helium contributes greatly to the appearance of the first adiabatic exponent ($\Gamma_1$) profile by shaping the size of the helium wells, the state of degeneracy seems to affect the profile just as much by altering their dispersion. Taking this into consideration, the model is thus able to describe a wide variety of ionisation structures while providing them with a physical meaning. By comparing them to a CESTAM model, we also verified that it was able to approach realistic $\Gamma_1$ profiles for consistent parameter values, conferring it a predictive capacity.

The modelling work conducted allowed us to study the shape of structural perturbations by analysing differences of profile with distinct parameterisations. In particular, the form of the perturbations caused by a helium difference was addressed. Expected shapes were found such as a Gaussian-like contribution for the helium ionisation regions with in addition a sharper component at the surface, caused by a shift in the hydrogen ionisation region. However, we have observed that this form is only valid for a restricted case, namely a helium difference under solar electron degeneracy conditions. This perturbation seems, however, highly variable with respect to these conditions, and can easily become much more complex at lower electron degeneracies. This point stresses, in particular, the major dependence of the perturbation on the choice of realistic models chosen for calibration purposes. We then went on to study more general parameter differences. More elaborate forms of perturbations than the ones usually assumed are found. It is also suggested that there is a stronger dependence on the electron degeneracy in the CZ or on the position of the ionisation region than on the helium amount itself. Also, the connection between electron degeneracy and stellar mass therefore enables us to clarify the strong dependence of the helium glitch amplitude on the stellar mass already observed by \citet{Verma2014b, Verma2019}. Moreover, the fluctuation in the ionisation region position thus induced seems compatible with a glitch signature coming from the peak of the first adiabatic exponent as reported by \cite{Broomhall2014} or \cite{Verma2014b}.

When analysing the oscillation frequencies, we therefore emphasise the importance of having a relationship that can take into account these additional dependencies.  In this sense, a second paper based on the introduced modelling is planned in order to derive more general forms of frequency shifts. The objective should be to interpret the ionisation glitch as a combination of multiple contributions for which a variation of helium abundance is one component among others.


%
%
\bibliographystyle{aa}
\bibliography{src.bib}

\begin{thebibliography}{53}
\expandafter\ifx\csname natexlab\endcsname\relax\def\natexlab#1{#1}\fi

\bibitem[{{Antia} \& {Basu}(1994)}]{Antia1994}
{Antia}, H.~M. \& {Basu}, S. 1994, \aaps, 107, 421

\bibitem[{{Asplund} {et~al.}(2009){Asplund}, {Grevesse}, {Sauval}, \&
  {Scott}}]{Asplund2009}
{Asplund}, M., {Grevesse}, N., {Sauval}, A.~J., \& {Scott}, P. 2009, \araa, 47,
  481

\bibitem[{{Baglin} {et~al.}(2006){Baglin}, {Auvergne}, {Barge}, {Deleuil},
  {Catala}, {Michel}, {Weiss}, \& {COROT Team}}]{Baglin2006}
{Baglin}, A., {Auvergne}, M., {Barge}, P., {et~al.} 2006, in ESA Special
  Publication, Vol. 1306, The CoRoT Mission Pre-Launch Status - Stellar
  Seismology and Planet Finding, ed. M.~{Fridlund}, A.~{Baglin}, J.~{Lochard},
  \& L.~{Conroy}, 33

\bibitem[{{Baker} \& {Kippenhahn}(1962)}]{Baker1962}
{Baker}, N. \& {Kippenhahn}, R. 1962, \zap, 54, 114

\bibitem[{{Ball} \& {Gizon}(2014)}]{Ball2014}
{Ball}, W.~H. \& {Gizon}, L. 2014, \aap, 568, A123

\bibitem[{{Ballot} {et~al.}(2004){Ballot}, {Turck-Chi{\`e}ze}, \&
  {Garc{\'\i}a}}]{Ballot2004}
{Ballot}, J., {Turck-Chi{\`e}ze}, S., \& {Garc{\'\i}a}, R.~A. 2004, \aap, 423,
  1051

\bibitem[{{Basu} \& {Antia}(2004)}]{Basu2004b}
{Basu}, S. \& {Antia}, H.~M. 2004, \apjl, 606, L85

\bibitem[{{Basu} {et~al.}(1994){Basu}, {Antia}, \& {Narasimha}}]{Basu1994}
{Basu}, S., {Antia}, H.~M., \& {Narasimha}, D. 1994, \mnras, 267, 209

\bibitem[{{Belkacem} {et~al.}(2021){Belkacem}, {Kupka}, {Philidet}, \&
  {Samadi}}]{Belkacem2021}
{Belkacem}, K., {Kupka}, F., {Philidet}, J., \& {Samadi}, R. 2021, \aap, 646,
  L5

\bibitem[{{Belkacem} {et~al.}(2019){Belkacem}, {Kupka}, {Samadi}, \&
  {Grimm-Strele}}]{Belkacem2019}
{Belkacem}, K., {Kupka}, F., {Samadi}, R., \& {Grimm-Strele}, H. 2019, \aap,
  625, A20

\bibitem[{{Broomhall} {et~al.}(2014){Broomhall}, {Miglio}, {Montalb{\'a}n},
  {Eggenberger}, {Chaplin}, {Elsworth}, {Scuflaire}, {Ventura}, \&
  {Verner}}]{Broomhall2014}
{Broomhall}, A.~M., {Miglio}, A., {Montalb{\'a}n}, J., {et~al.} 2014, MNRAS,
  440, 1828

\bibitem[{{Canuto}(1997)}]{Canuto1997}
{Canuto}, V.~M. 1997, \apj, 482, 827

\bibitem[{{Christensen-Dalsgaard} {et~al.}(1988){Christensen-Dalsgaard},
  {Dappen}, \& {Lebreton}}]{JCD1988}
{Christensen-Dalsgaard}, J., {Dappen}, W., \& {Lebreton}, Y. 1988, \nat, 336,
  634

\bibitem[{{Christensen-Dalsgaard} {et~al.}(1993){Christensen-Dalsgaard},
  {Proffitt}, \& {Thompson}}]{JCD1993}
{Christensen-Dalsgaard}, J., {Proffitt}, C.~R., \& {Thompson}, M.~J. 1993,
  \apjl, 403, L75

\bibitem[{{Christensen-Dalsgaard} \& {Thompson}(1997)}]{JCD1997}
{Christensen-Dalsgaard}, J. \& {Thompson}, M.~J. 1997, \mnras, 284, 527

\bibitem[{{Cox} \& {Giuli}(1968)}]{Cox1968}
{Cox}, J.~P. \& {Giuli}, R.~T. 1968, {Principles of stellar structure}

\bibitem[{{Deal} {et~al.}(2018){Deal}, {Alecian}, {Lebreton}, {Goupil},
  {Marques}, {LeBlanc}, {Morel}, \& {Pichon}}]{Deal2018}
{Deal}, M., {Alecian}, G., {Lebreton}, Y., {et~al.} 2018, A\&A, 618, A10

\bibitem[{{Dziembowski} {et~al.}(1990){Dziembowski}, {Pamyatnykh}, \&
  {Sienkiewicz}}]{Dziembowski1990}
{Dziembowski}, W.~A., {Pamyatnykh}, A.~A., \& {Sienkiewicz}, R. 1990, MNRAS,
  244, 542

\bibitem[{{Farnir} {et~al.}(2019){Farnir}, {Dupret}, {Salmon}, {Noels}, \&
  {Buldgen}}]{Farnir2019}
{Farnir}, M., {Dupret}, M.~A., {Salmon}, S.~J.~A.~J., {Noels}, A., \&
  {Buldgen}, G. 2019, \aap, 622, A98

\bibitem[{{Gilliland} {et~al.}(2010){Gilliland}, {Brown},
  {Christensen-Dalsgaard}, {Kjeldsen}, {Aerts}, {Appourchaux}, {Basu},
  {Bedding}, {Chaplin}, {Cunha}, {De Cat}, {De Ridder}, {Guzik}, {Handler},
  {Kawaler}, {Kiss}, {Kolenberg}, {Kurtz}, {Metcalfe}, {Monteiro}, {Szab{\'o}},
  {Arentoft}, {Balona}, {Debosscher}, {Elsworth}, {Quirion}, {Stello},
  {Su{\'a}rez}, {Borucki}, {Jenkins}, {Koch}, {Kondo}, {Latham}, {Rowe}, \&
  {Steffen}}]{Gilliland2010}
{Gilliland}, R.~L., {Brown}, T.~M., {Christensen-Dalsgaard}, J., {et~al.} 2010,
  \pasp, 122, 131

\bibitem[{{Gough}(1990)}]{Gough1990}
{Gough}, D.~O. 1990, in Progress of Seismology of the Sun and Stars,
  Proceedings of the Oji International Seminar Held at Hakone, Japan, 11-14
  December 1989. Lecture Notes in Physics, Vol. 367, edited by Y. Osaki and H.
  Shibahashi. Springer-Verlag, Berlin Heidelberg New York, 1990, p.283, ed.
  Y.~{Osaki} \& H.~{Shibahashi}, Vol. 367, 283

\bibitem[{{Gough}(2002)}]{Gough2002}
{Gough}, D.~O. 2002, in ESA Special Publication, Vol. 485, Stellar Structure
  and Habitable Planet Finding, ed. B.~{Battrick}, F.~{Favata}, I.~W.
  {Roxburgh}, \& D.~{Galadi}, 65--73

\bibitem[{{Gough} \& {Thompson}(1991)}]{Gough1991}
{Gough}, D.~O. \& {Thompson}, M.~J. 1991, {The inversion problem} (University
  of Arizona Press, Tucson), 519--561

\bibitem[{{Hayashi} {et~al.}(1962){Hayashi}, {H{\={o}}shi}, \&
  {Sugimoto}}]{Hayashi1962}
{Hayashi}, C., {H{\={o}}shi}, R., \& {Sugimoto}, D. 1962, Progress of
  Theoretical Physics Supplement, 22, 1

\bibitem[{{Houdek} \& {Gough}(2007)}]{Houdek2007}
{Houdek}, G. \& {Gough}, D.~O. 2007, MNRAS, 375, 861

\bibitem[{{Houdek} \& {Gough}(2011)}]{Houdek2011}
{Houdek}, G. \& {Gough}, D.~O. 2011, \mnras, 418, 1217

\bibitem[{{Kippenhahn} {et~al.}(2012){Kippenhahn}, {Weigert}, \&
  {Weiss}}]{Kippenhahn2012}
{Kippenhahn}, R., {Weigert}, A., \& {Weiss}, A. 2012, {Stellar Structure and
  Evolution} (Berlin Heidelberg: Springer-Verlag, 2nd Edition)

\bibitem[{{Kjeldsen} {et~al.}(2008){Kjeldsen}, {Bedding}, \&
  {Christensen-Dalsgaard}}]{Kjeldsen2008}
{Kjeldsen}, H., {Bedding}, T.~R., \& {Christensen-Dalsgaard}, J. 2008, \apjl,
  683, L175

\bibitem[{{Kosovichev} {et~al.}(1992){Kosovichev}, {Christensen-Dalsgaard},
  {Daeppen}, {Dziembowski}, {Gough}, \& {Thompson}}]{Kosovichev1992}
{Kosovichev}, A.~G., {Christensen-Dalsgaard}, J., {Daeppen}, W., {et~al.} 1992,
  \mnras, 259, 536

\bibitem[{{Lebreton} \& {Goupil}(2014)}]{Lebreton2014}
{Lebreton}, Y. \& {Goupil}, M.~J. 2014, A\&A, 569, A21

\bibitem[{Lopes {et~al.}(1997)Lopes, Turck-Chieze, Michel, \&
  Goupil}]{Lopes1997}
Lopes, I., Turck-Chieze, S., Michel, E., \& Goupil, M.-J. 1997, The
  Astrophysical Journal, 480, 794

\bibitem[{{Lund} {et~al.}(2017){Lund}, {Silva Aguirre}, {Davies}, {Chaplin},
  {Christensen-Dalsgaard}, {Houdek}, {White}, {Bedding}, {Ball}, {Huber},
  {Antia}, {Lebreton}, {Latham}, {Handberg}, {Verma}, {Basu}, {Casagrande},
  {Justesen}, {Kjeldsen}, \& {Mosumgaard}}]{Lund2017}
{Lund}, M.~N., {Silva Aguirre}, V., {Davies}, G.~R., {et~al.} 2017, \apj, 835,
  172

\bibitem[{{Marques} {et~al.}(2013){Marques}, {Goupil}, {Lebreton}, {Talon},
  {Palacios}, {Belkacem}, {Ouazzani}, {Mosser}, {Moya}, {Morel}, {Pichon},
  {Mathis}, {Zahn}, {Turck-Chi{\`e}ze}, \& {Nghiem}}]{Marques2013}
{Marques}, J.~P., {Goupil}, M.~J., {Lebreton}, Y., {et~al.} 2013, \aap, 549,
  A74

\bibitem[{{Monteiro} {et~al.}(1994){Monteiro}, {Christensen-Dalsgaard}, \&
  {Thompson}}]{Monteiro1994}
{Monteiro}, M.~J.~P.~F.~G., {Christensen-Dalsgaard}, J., \& {Thompson}, M.~J.
  1994, A\&A, 283, 247

\bibitem[{{Monteiro} \& {Thompson}(1998)}]{Monteiro1998}
{Monteiro}, M.~J.~P.~F.~G. \& {Thompson}, M.~J. 1998, in IAU Symposium, Vol.
  185, New Eyes to See Inside the Sun and Stars, ed. F.-L. {Deubner},
  J.~{Christensen-Dalsgaard}, \& D.~{Kurtz}, 317

\bibitem[{{Monteiro} \& {Thompson}(2005)}]{Monteiro2005}
{Monteiro}, M. J.~P.~F.~G. \& {Thompson}, M.~J. 2005, MNRAS, 361, 1187

\bibitem[{{Morel} \& {Lebreton}(2008)}]{Morel2008}
{Morel}, P. \& {Lebreton}, Y. 2008, \apss, 316, 61

\bibitem[{{Mosumgaard} {et~al.}(2020){Mosumgaard}, {J{\o}rgensen}, {Weiss},
  {Silva Aguirre}, \& {Christensen-Dalsgaard}}]{Mosumgaard2020}
{Mosumgaard}, J.~R., {J{\o}rgensen}, A. C.~S., {Weiss}, A., {Silva Aguirre},
  V., \& {Christensen-Dalsgaard}, J. 2020, \mnras, 491, 1160

\bibitem[{{Noll} {et~al.}(2021){Noll}, {Deheuvels}, \& {Ballot}}]{Noll2021}
{Noll}, A., {Deheuvels}, S., \& {Ballot}, J. 2021, \aap, 647, A187

\bibitem[{{Nsamba} {et~al.}(2018){Nsamba}, {Campante}, {Monteiro}, {Cunha},
  {Rendle}, {Reese}, \& {Verma}}]{Nsamba2018}
{Nsamba}, B., {Campante}, T.~L., {Monteiro}, M.~J.~P.~F.~G., {et~al.} 2018,
  \mnras, 477, 5052

\bibitem[{{Perez Hernandez} \&
  {Christensen-Dalsgaard}(1994)}]{PerezHernandez1994}
{Perez Hernandez}, F. \& {Christensen-Dalsgaard}, J. 1994, \mnras, 269, 475

\bibitem[{{Ricker} {et~al.}(2015){Ricker}, {Winn}, {Vanderspek}, {Latham},
  {Bakos}, {Bean}, {Berta-Thompson}, {Brown}, {Buchhave}, {Butler}, {Butler},
  {Chaplin}, {Charbonneau}, {Christensen-Dalsgaard}, {Clampin}, {Deming},
  {Doty}, {De Lee}, {Dressing}, {Dunham}, {Endl}, {Fressin}, {Ge}, {Henning},
  {Holman}, {Howard}, {Ida}, {Jenkins}, {Jernigan}, {Johnson}, {Kaltenegger},
  {Kawai}, {Kjeldsen}, {Laughlin}, {Levine}, {Lin}, {Lissauer}, {MacQueen},
  {Marcy}, {McCullough}, {Morton}, {Narita}, {Paegert}, {Palle}, {Pepe},
  {Pepper}, {Quirrenbach}, {Rinehart}, {Sasselov}, {Sato}, {Seager},
  {Sozzetti}, {Stassun}, {Sullivan}, {Szentgyorgyi}, {Torres}, {Udry}, \&
  {Villasenor}}]{Ricker2015}
{Ricker}, G.~R., {Winn}, J.~N., {Vanderspek}, R., {et~al.} 2015, Journal of
  Astronomical Telescopes, Instruments, and Systems, 1, 014003

\bibitem[{{Rogers}(1981)}]{Rogers1981}
{Rogers}, F.~J. 1981, \pra, 24, 1531

\bibitem[{{Rogers} {et~al.}(1996){Rogers}, {Swenson}, \&
  {Iglesias}}]{Rogers1996}
{Rogers}, F.~J., {Swenson}, F.~J., \& {Iglesias}, C.~A. 1996, \apj, 456, 902

\bibitem[{{Schou} \& {Birch}(2020)}]{Schou2020}
{Schou}, J. \& {Birch}, A.~C. 2020, \aap, 638, A51

\bibitem[{{Serenelli} {et~al.}(2009){Serenelli}, {Basu}, {Ferguson}, \&
  {Asplund}}]{Serenelli2009}
{Serenelli}, A.~M., {Basu}, S., {Ferguson}, J.~W., \& {Asplund}, M. 2009,
  \apjl, 705, L123

\bibitem[{{Sonoi} {et~al.}(2015){Sonoi}, {Samadi}, {Belkacem}, {Ludwig},
  {Caffau}, \& {Mosser}}]{Sonoi2015}
{Sonoi}, T., {Samadi}, R., {Belkacem}, K., {et~al.} 2015, \aap, 583, A112

\bibitem[{{Stassun} {et~al.}(2019){Stassun}, {Oelkers}, {Paegert}, {Torres},
  {Pepper}, {De Lee}, {Collins}, {Latham}, {Muirhead}, {Chittidi},
  {Rojas-Ayala}, {Fleming}, {Rose}, {Tenenbaum}, {Ting}, {Kane}, {Barclay},
  {Bean}, {Brassuer}, {Charbonneau}, {Ge}, {Lissauer}, {Mann}, {McLean},
  {Mullally}, {Narita}, {Plavchan}, {Ricker}, {Sasselov}, {Seager}, {Sharma},
  {Shiao}, {Sozzetti}, {Stello}, {Vanderspek}, {Wallace}, \&
  {Winn}}]{Stassun2019}
{Stassun}, K.~G., {Oelkers}, R.~J., {Paegert}, M., {et~al.} 2019, \aj, 158, 138

\bibitem[{{Tassoul}(1980)}]{Tassoul1980}
{Tassoul}, M. 1980, ApJS, 43, 469

\bibitem[{{Verma} {et~al.}(2014{\natexlab{a}}){Verma}, {\GG{a}}{Faria},
  {Antia}, {Basu}, {Mazumdar}, {Monteiro}, {Appourchaux}, {Chaplin},
  {Garc{\'\i}a}, \& {Metcalfe}}]{Verma2014a}
{Verma}, K., {\GG{a}}{Faria}, J.~P., {Antia}, H.~M., {et~al.}
  2014{\natexlab{a}}, \apj, 790, 138

\bibitem[{{Verma} {et~al.}(2014{\natexlab{b}}){Verma}, {\GG{b}}{Antia}, {Basu},
  \& {Mazumdar}}]{Verma2014b}
{Verma}, K., {\GG{b}}{Antia}, H.~M., {Basu}, S., \& {Mazumdar}, A.
  2014{\natexlab{b}}, \apj, 794, 114

\bibitem[{{Verma} {et~al.}(2017){Verma}, {Raodeo}, {Antia}, {Mazumdar}, {Basu},
  {Lund}, \& {Silva Aguirre}}]{Verma2017}
{Verma}, K., {Raodeo}, K., {Antia}, H.~M., {et~al.} 2017, \apj, 837, 47

\bibitem[{{Verma} {et~al.}(2019){Verma}, {Raodeo}, {Basu}, {Silva Aguirre},
  {Mazumdar}, {Mosumgaard}, {Lund}, \& {Ranadive}}]{Verma2019}
{Verma}, K., {Raodeo}, K., {Basu}, S., {et~al.} 2019, \mnras, 483, 4678

\end{thebibliography}

\appendix

\section{Further simplifications and perturbation of the first adiabatic exponent \label{A}}

\subsection{Deriving Eq. \eqref{eq:G1g1}}

This parts details the derivation of \eqref{eq:G1g1} and \eqref{eq:g1} from \eqref{eq:G1d2f}-\eqref{eq:dTTf}. Let us first introduce the notation:
\begin{equation}
    \label{eq:Chindef}
    \chi_n \equiv \sum_{ir}x_iy_i^r(1-y_i^r)\frac{{(\chi_i^r/kT)}^{n}}{1+\delta_i^1(1-y_i^r)}
\end{equation}
Equations \eqref{eq:dVVf}-\eqref{eq:dTTf} can then be rewritten as:
\begin{align}
    \partial_{VV}^2 f ~&=~ \partial_{V} f-\chi_0\\
    \partial_{TV}^2 f = \partial_{VT}^2 f ~&=~ \partial_{V} f + \chi_1 + \frac{3}{2}\chi_0\\
    \label{eq:dTTfChi}
    \partial_{TT}^2 f ~&=~ \frac{3}{2}\partial_{V} f + \chi_2 + 3\chi_1 +\frac{9}{4}\chi_0
\end{align}

Injecting this into Eq. \eqref{eq:G1d2f} then leads to:
\begin{equation}
    \Gamma_1 = \frac{\left(\partial_{V} f + \chi_1 + \frac{3}{2}\chi_0\right)^2+\left(\partial_{V} f-\chi_0\right)\left(\frac{3}{2}\partial_{V} f + \chi_2 + 3\chi_1 +\frac{9}{4}\chi_0\right)}{\partial_{V} f\left(\frac{3}{2}\partial_{V} f + \chi_2 + 3\chi_1 +\frac{9}{4}\chi_0\right)}
\end{equation}and since:
\begin{equation}
    \begin{split}
        \left(\partial_{V} f + \chi_1 + \frac{3}{2}\chi_0\right)^2 &= \partial_{V} f\left(\partial_{V} f + 2\chi_1 + 3\chi_0\right) +\left(\chi_1 + \frac{3}{2}\chi_0\right)^2 \\
        &= \frac{2}{3}\partial_{V} f\left(\frac{3}{2}\partial_{V} f + \chi_2 + 3\chi_1 +\frac{9}{4}\chi_0\right) \\
        &~~~+ \partial_{V} f\left(-\frac{2}{3}\chi_2 + \frac{3}{2}\chi_0\right) +\left(\chi_1 + \frac{3}{2}\chi_0\right)^2
    \end{split}
\end{equation}we obtain the following expression in which $5/3$ appears as expected:
\begin{equation}
    \begin{split}
        \Gamma_1 =\frac{5}{3}~&-~\frac{\partial_{V} f\left(\frac{2}{3}\chi_2 - \frac{3}{2}\chi_0\right)-\left(\chi_1 + \frac{3}{2}\chi_0\right)^2}{\partial_{V} f\left(\frac{3}{2}\partial_{V} f + \chi_2 + 3\chi_1 +\frac{9}{4}\chi_0\right)}\\
        ~&+~\frac{\chi_0\left(\frac{3}{2}\partial_{V} f + \chi_2 + 3\chi_1 +\frac{9}{4}\chi_0\right)}{\partial_{V} f\left(\frac{3}{2}\partial_{V} f + \chi_2 + 3\chi_1 +\frac{9}{4}\chi_0\right)}
    \end{split}
\end{equation}

The simplifications in the numerator lead to the following more compact expression:
\begin{equation}
    \Gamma_1 ~=~\frac{5}{3}-\frac{\partial_{V} f\left(\frac{2}{3}\chi_2\right)+\chi_2\chi_0-\left(\chi_1\right)^2}{\partial_{V} f\left(\frac{3}{2}\partial_{V} f + \chi_2 + 3\chi_1 +\frac{9}{4}\chi_0\right)}
\end{equation}

Let us look at the term $\chi_2\chi_0-\left(\chi_1\right)^2$:
\begin{equation}
    \begin{split}
        \chi_2\chi_0-\left(\chi_1\right)^2 &= \sum_{ijrs}x_ix_jy_i^ry_j^s\frac{(1-y_i^r)(1-y_j^s)\chi_i^r(\chi_i^r-\chi_j^s)/(kT)^2}{\left[1+\delta_i^1(1-y_i^r)\right]\left[1+\delta_j^1(1-y_j^s)\right]} \\
        &= \sum_{ir<js}x_ix_jy_i^ry_j^s\frac{(1-y_i^r)(1-y_j^s)(\chi_i^r-\chi_j^s)^2/(kT)^2}{\left[1+\delta_i^1(1-y_i^r)\right]\left[1+\delta_j^1(1-y_j^s)\right]}\\
        &< \sum_{ir<js}x_ix_jy_i^ry_j^s(1-y_i^r)(1-y_j^s)\left(\frac{\chi_i^r-\chi_j^s}{kT}\right)^2
    \end{split}
\end{equation}where we grouped the symmetrical terms under the notation ``$ir<js$''. One may note that most of the terms in this sum are negligible compared with the denominator which is always greater than $3/2$. There are several effects that contribute to keeping these terms small:
\begin{enumerate}
    \item In order for the product $x_i x_j$ to contribute to the total, the number abundances of both elements, $i$ and $j$, need to be non-negligible.  Accordingly, the only non-negligible terms are those coupled to hydrogen.
    \item The ionisation regions $(i,r)$ and $(j,s)$ should overlap as much as possible since otherwise $y_i^ry_j^s(1-y_i^r)(1-y_j^s)$ would be null. Even in the most favourable case, we have the upper limit $y_i^ry_j^s(1-y_i^r)(1-y_j^s) < 1/16$.
    \item At the same time, ionisation regions that overlap tend to have a similar ionisation potential, thus making $(\chi_i^r-\chi_j^s)^2$ small.
    \item As discussed in Section \ref{MOD}, ionisation regions are more likely to merge at higher conditions of temperature and density. However in this case, it is $1/(kT)^2$ that tends rapidly towards $0$. 
\end{enumerate}

All these points seem to be arguing in favour of a small $\chi_2\chi_0-\left(\chi_1\right)^2$ and we confirm numerically that this term remains below $2\times10^{-3}$ at any position of the $(\rho,T)$ plane. We thus considered that $\chi_2\chi_0-\left(\chi_1\right)^2 = 0$ for the remainder of our derivation. Eq. \eqref{eq:G1g1} quickly ensues by introducing:
\begin{equation}
    \label{eq:g1A}
    \gamma_1 \equiv \frac{\chi_2}{\frac{3}{2}\partial_{V} f + \chi_2 + 3\chi_1 +\frac{9}{4}\chi_0} = \frac{\chi_2}{\partial_{TT}^2 f}
\end{equation}
Relation \eqref{eq:g1} is simply obtained by replacing $\chi_2$ by its definition \eqref{eq:Chindef}. As can be seen from Eq.~\eqref{eq:dTTfChi}, $\partial_{TT}^2 f > \chi_2$ and this results in the property stated in the main part, namely $0\leq \gamma_1 < 1$, or equivalently $1<\Gamma_1\leq 5/3$.

\subsection{Perturbing $\Gamma_1$ at given $V,T$ values}

The resulting relation, Eq. \eqref{eq:g1A}, is so compact that we can perturb it analytically in terms of abundances (which was not an option when considering Eq.~\eqref{eq:G1d2f}). Let us consider a change $\delta x_i$ of the $x_i$ at given $V$ and $T$ values (with the implicit condition $\sum_i \delta x_i = 0$). Considering Eqs.~\eqref{eq:Saha5}-\eqref{eq:yir}, the $y_i^r$ thus defined are functions of $T$ and $V$ but not of $x_i$. It can thus be established that $\delta_{V,T}(y_i^r) = 0$. The induced perturbations of $\chi_2$ and $\partial_{TT}^2f$ will therefore be given by:
\begin{align}
    \delta_{V,T}(\chi_2) &= \sum_{ir} \frac{\delta x_i}{x_i} x_i y_i^r(1-y_i^r)\frac{{(\chi_i^r/kT)}^{2}}{1+\delta_i^1(1-y_i^r)} \\
    \delta_{V,T}(\partial_{TT}^2f) &= \sum_{ir}\frac{\delta x_i}{x_i} x_i y_i^r\left(\frac{3}{2}+(1-y_i^r)\frac{{(3/2+\chi_i^r/kT)}^{2}}{1+\delta_i^1(1-y_i^r)}\right)
\end{align}thus leading to the following impact on $\gamma_1$:
\begin{equation}
    \delta_{V,T}(\gamma_1) = \frac{(\partial_{TT}^2f)\delta_{V,T}(\chi_2)-(\chi_2)\delta_{V,T}(\partial_{TT}^2f)}{(\partial_{TT}^2f)^2}
\end{equation}

By calculating the numerator explicitly, one can see the apparition of 4 sums:
\begin{equation}
    \begin{split}
        (\partial_{TT}^2f)^2 \delta_{V,T}(\gamma_1) ~&=~ \frac{3}{2}\sum_{ir} \frac{\delta x_i}{x_i} x_i y_i^r(1-y_i^r)\frac{{(\chi_i^r/kT)}^{2}}{1+\delta_i^1(1-y_i^r)}\\
        + \frac{3}{2}&\sum_{ijrs}\left(\frac{\delta x_i}{x_i}-\frac{\delta x_j}{x_j}\right)x_ix_jy_i^ry_j^s\frac{(1-y_i^r)(\chi_i^r/kT)^2}{1+\delta_i^1(1-y_i^r)}\\
        + \frac{9}{4}\sum_{ijrs} \frac{\delta x_i}{x_i}&x_ix_jy_i^ry_j^s\frac{(1-y_i^r)(1-y_j^s)(\chi_i^r+\chi_j^s)(\chi_i^r-\chi_j^s)/(kT)^2}{\left[1+\delta_i^1(1-y_i^r)\right]\left[1+\delta_j^1(1-y_j^s)\right]}\\
        + 3\sum_{ijrs} \frac{\delta x_i}{x_i}&x_ix_jy_i^ry_j^s\frac{(1-y_i^r)(1-y_j^s)\chi_i^r\chi_j^s(\chi_i^r-\chi_j^s)/(kT)^3}{\left[1+\delta_i^1(1-y_i^r)\right]\left[1+\delta_j^1(1-y_j^s)\right]}
    \end{split}
\end{equation} If we group symmetrical terms, we see that the last two sums are bounded by:
\begin{equation}
    \begin{split}
        3^{\textrm{rd}}~\textrm{sum} < \frac{9}{4}\sum_{ir<js} \left(\frac{\delta x_i}{x_i}-\frac{\delta x_j}{x_j}\right)&x_ix_jy_i^ry_j^s(1-y_i^r)(1-y_j^s)\\
        &\times\left(\frac{\chi_i^r+\chi_j^s}{kT}\right)\left(\frac{\chi_i^r-\chi_j^s}{kT}\right)
    \end{split}
\end{equation}and
\begin{equation}
    \begin{split}
        4^{\textrm{th}}~\textrm{sum} < 3\sum_{ir<js} \left(\frac{\delta x_i}{x_i}-\frac{\delta x_j}{x_j}\right)&x_ix_jy_i^ry_j^s(1-y_i^r)(1-y_j^s)\\
        &\times \left(\frac{\chi_i^r}{kT}\right)\left(\frac{\chi_j^s}{kT}\right)\left(\frac{\chi_i^r-\chi_j^s}{kT}\right)
    \end{split}
\end{equation}

For the same reasons as mentioned above, we will consider that these two terms are negligible in the following. Since $\delta_{V,T}(\Gamma_1) = -2/3\delta_{V,T}(\gamma_1)$, an analytical expression of the perturbation is given by:
\begin{equation}
    \label{eq:dVTG1}
    \begin{split}
        \delta_{V,T}(\Gamma_1) ~\simeq~ -\frac{1}{(\partial_{TT}^2f)^2}&\sum_{ir} x_i y_i^r\frac{(1-y_i^r){(\chi_i^r/kT)}^{2}}{1+\delta_i^1(1-y_i^r)}\\
        &~~~~~\left[\frac{\delta x_i}{x_i}+\sum_{js}x_jy_j^s\left(\frac{\delta x_i}{x_i}-\frac{\delta x_j}{x_j}\right)\right]
    \end{split}
\end{equation}

\subsection{Perturbing $\Gamma_1$ at given $\rho,T$ values}

The perturbation at given $\rho$ and $T$ values is more complex since $\delta_{\rho,T}y_i^r \neq 0$ because of the variable change $\rho/m_0 = N/V$. For any quantity $\alpha$, the resulting perturbation change can be found by applying:
\begin{equation}
    \begin{split}
        \delta_{\rho,T}(\alpha) &= \delta_{V,T}(\alpha) - \delta_{V,T}(\rho) \left|\frac{\partial(\alpha,T)}{\partial(\rho,T)}\right| \\
        &= \delta_{V,T}(\alpha) - \frac{\delta m_0}{m_0}\left(\frac{\partial \alpha}{\partial \ln \rho}\right)_T
    \end{split}
\end{equation}

One can now find the perturbation of any quantity expressed in terms of $y_i^r$ using $\displaystyle \frac{\partial\ln y_i^r}{\partial\ln \rho} = \frac{d\ln y_i^r}{d\ln K_i^r}\frac{\partial\ln K_i^r}{\partial\ln \rho} = -\frac{1-y_i^r}{1+\delta_i^1(1-y_i^r)}$. To simplify the derivation, we will just present the parts that do not result in terms containing $x_ix_jy_i^ry_j^s(1-y_i^r)(1-y_j^s)$ which will be neglected:
\begin{equation}
    \begin{split}
        \delta_{\rho,T}(\chi_2) = \delta_{V,T}(\chi_2)+\frac{\delta m_0}{m_0}\sum_{ir} x_i y_i^r &\left(\frac{(1-y_i^r)(\chi_i^r/kT)}{1+\delta_i^1(1-y_i^r)}\right)^2\\
        &\left(1-\frac{y_i^r}{1-y_i^r}+\frac{\delta_i^1y_i^r}{1+\delta_i^1(1-y_i^r)}\right)
    \end{split}
\end{equation} The resulting perturbation of the numerator is approximated by:
\begin{equation}
    \begin{split}
        (\partial_{TT}^2f)^2 \delta_{\rho,T}(\gamma_1) ~&=~ (\partial_{TT}^2f)^2 \delta_{V,T}(\gamma_1) + (\partial_{TT}^2f)^2\left[\delta_{\rho,T}(\chi_2)-\delta_{V,T}(\chi_2)\right] \\
        &~~~~~~~~~~~~~~~~~~~~~~~~~~~~~~~- (\chi_2)\left[\delta_{\rho,T}(\partial_{TT}^2f)-\delta_{V,T}(\partial_{TT}^2f)\right] \\
        ~&\simeq~ (\partial_{TT}^2f)^2 \delta_{V,T}(\gamma_1) + \frac{3}{2}(\partial_V f)\left[\delta_{\rho,T}(\chi_2)-\delta_{V,T}(\chi_2)\right] \\
        ~&\simeq~ (\partial_{TT}^2f)^2 \delta_{V,T}(\gamma_1) \\
        + \frac{3}{2}\frac{\delta m_0}{m_0}&\sum_{ir} x_i y_i^r\left(\frac{(1-y_i^r)(\chi_i^r/kT)}{1+\delta_i^1(1-y_i^r)}\right)^2\\
        &~~~~~~\left(\frac{2-y_i^r}{1-y_i^r}+\frac{\delta_i^1y_i^r}{1+\delta_i^1(1-y_i^r)}\right)\left(1+\sum_{js}x_jy_j^s\right)
    \end{split}
\end{equation}

And finally the overall perturbation on $\delta_{\rho,T}(\Gamma_1)$ will thus be given by the two sums:
\begin{equation}
    \label{eq:drhoTG1}
    \begin{split}
        \delta_{\rho,T}(\Gamma_1) ~\simeq~ -\frac{1}{(\partial_{TT}^2f)^2}&\sum_{ir} x_i y_i^r\frac{(1-y_i^r){(\chi_i^r/kT)}^{2}}{1+\delta_i^1(1-y_i^r)}\\
        &~~~~~\left[\frac{\delta x_i}{x_i}+\sum_{js}x_jy_j^s\left(\frac{\delta x_i}{x_i}-\frac{\delta x_j}{x_j}\right)\right]\\
        -\frac{\delta m_0/m_0}{(\partial_{TT}^2f)^2}&\sum_{ir} x_i y_i^r\left(\frac{(1-y_i^r)(\chi_i^r/kT)}{1+\delta_i^1(1-y_i^r)}\right)^2\\
         &~~~~~~\left(\frac{2-y_i^r}{1-y_i^r}+\frac{\delta_i^1y_i^r}{1+\delta_i^1(1-y_i^r)}\right)\left(1+\sum_{js}x_jy_j^s\right)
    \end{split}
\end{equation}

\begin{figure}
   \centering
   \includegraphics[width=9cm]{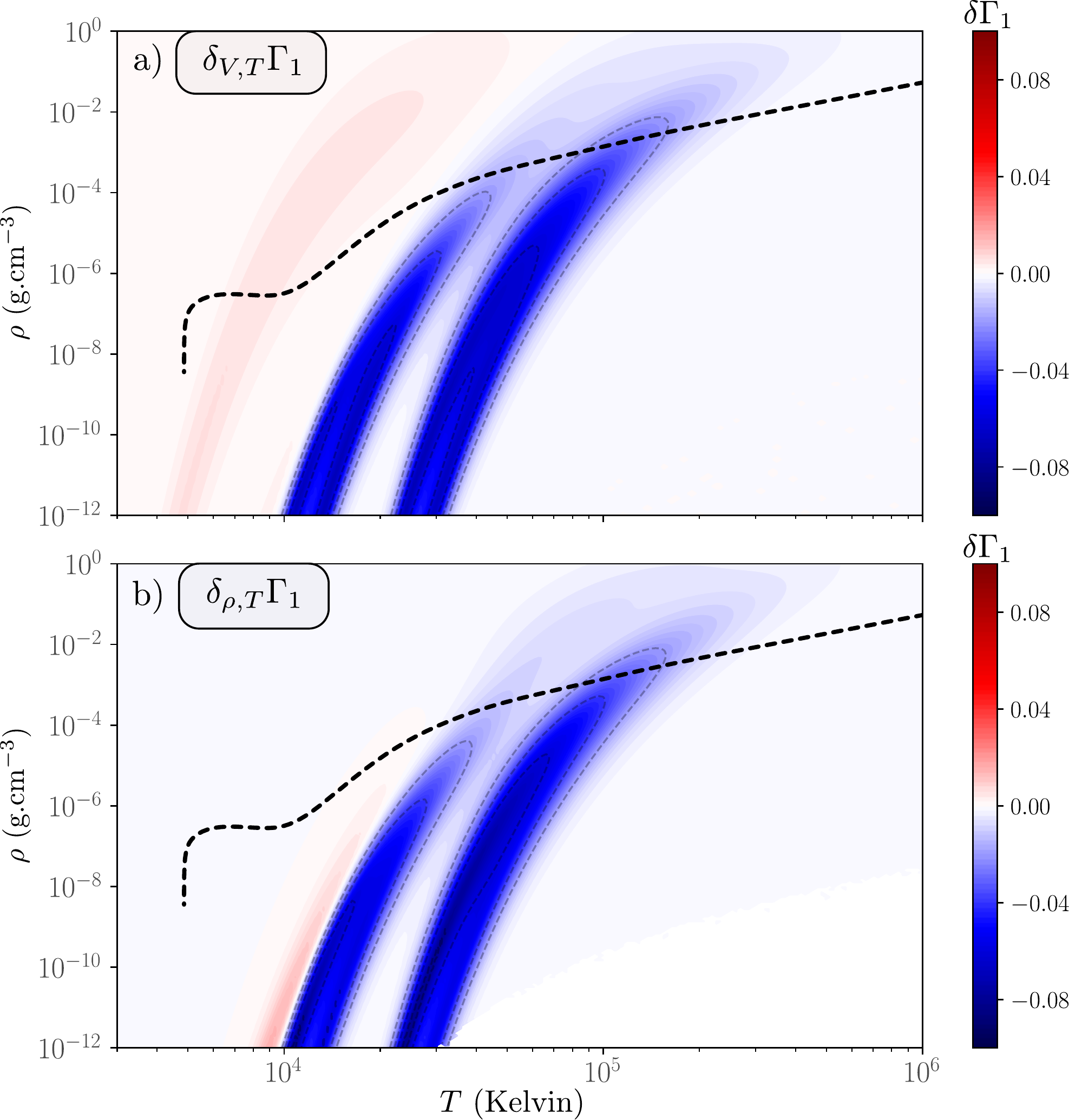}
      \caption{Representation of the approximations \eqref{eq:dVTG1} (\textit{panel (a)}) and \eqref{eq:drhoTG1} (\textit{panel (b)}) given by a change $\delta Y = 0.1$ from the reference value $Y=0.25$.}
         \label{MapdG1.pdf}
   \end{figure}

Although we have neglected many terms to obtain Eq.~\eqref{eq:drhoTG1}, this equation still gives a useful approximation. For instance, the difference between $\Gamma_1(\rho,T,0.35)-\Gamma_1(\rho,T,0.25)$ (shown in Fig. \ref{MapG1.pdf}) and the relation \eqref{eq:drhoTG1} is bounded by $10^{-2}$ at any position of the $(\rho,T)$ plane and one can verify that the panel (b) of Fig. \ref{MapdG1.pdf} is actually very similar to what has been found in the panel (d) of Fig. \ref{MapG1.pdf}. We have also represented the perturbation given by \eqref{eq:dVTG1} which seems closer to the shape that can be expected for the perturbation, at least based on the panel (b) of Figure \ref{MapG1.pdf}. The fact that the perturbation moves towards the borders of each ionisation region when passing from $\delta_{V,T}(\Gamma_1)$ to $\delta_{\rho,T}(\Gamma_1)$ is then interpreted as the result of a shift in the $y_i^r$ caused by perturbing at constant $\rho$. In other words, at a given density, a perturbation on the abundances makes the ionisation regions move and this results in a less intuitive map. This effect can however be seen as ``artificial'' since it is obtained by fixing a variable that mixes the specific effects of various state variables (i.e. $V$ and each $N_i$).

To give a qualitative understanding of the amplitude difference between the modification of the hydrogen and helium wells visible on the top panel of Fig. \ref{MapdG1.pdf}, one may consider the dominant term of the denominator in Eq.~\eqref{eq:dVTG1}:
\begin{equation}
    (\partial_{TT}^2f)^2 \underset{ir}{\sim} 3x_i y_i^r\frac{(1-y_i^r){(\chi_i^r/kT)}^{2}}{1+\delta_i^1(1-y_i^r)}\left(1+\sum_{js}x_jy_j^s\right)
\end{equation}where the ``$\underset{ir}{\sim}$'' notation stands for the dominant term in the ionisation region $(i,r)$. Comparing this with the numerator of Eq.~\eqref{eq:dVTG1} gives an intuition of why the order of magnitude to keep in mind for the perturbation is $\delta x_i/x_i$ rather than $\delta x_i$. Thus, while we have $\delta x_1 = -\delta x_2$ for the hydrogen-helium case, we see that $\delta x_1 / x_1 = -\delta x_2/x_1 = - (x_2/x_1) \delta x_2 / x_2$. However for $Y=0.25$, $(x_2/x_1) = 1/12$, resulting in a perturbation about $12$ times lower in the hydrogen region. This estimate is consistent with Fig.~\ref{MapdG1.pdf}, where the maximum in the hydrogen region is $\sim 6\times10^{-3}$ and the minimum in the helium region $\sim -7.2\times 10^{-2}$.
\end{document}